\documentclass[letterpaper,twocolumn,10pt]{article}
\usepackage{usenix2019_v3}

\usepackage{cite}
\usepackage{amsmath,amssymb,amsfonts}
\usepackage{algorithmic}
\usepackage{graphicx}
\usepackage{textcomp}
\usepackage{xcolor}
\usepackage{subfigure}
\usepackage[ruled, linesnumbered]{algorithm2e}
\usepackage{multirow}
\usepackage{booktabs}
\usepackage{epsfig,endnotes}
\usepackage{mathtools}
\usepackage{url}
\usepackage{xcolor}
\usepackage{epstopdf}
\usepackage{array}
\usepackage{color}
\usepackage[numbers]{natbib}
\usepackage{pifont}
\usepackage{tabularx}
\usepackage{bbm}
\usepackage{bbding}
\usepackage{gensymb}
\usepackage{diagbox}
\usepackage{colortbl}
\usepackage{float}
\usepackage[T1]{fontenc}
\usepackage{aecompl}
\usepackage[symbol]{footmisc}

\graphicspath{{figures/}}

\newcommand{\tp}{\texttt{TPatch}\xspace}

\newcommand{\tpes}{\texttt{TPatches}\xspace}

\begin{document}

\title{\Large \bf TPatch: A Triggered Physical Adversarial Patch}

\author{
{\rm Wenjun Zhu}\\
USSLAB, Zhejiang University \\
zwj\_@zju.edu.cn
\and
{\rm Xiaoyu Ji}\\
USSLAB, Zhejiang University \\
xji@zju.edu.cn
\and
{\rm Yushi Cheng*}\\
BNRist, Tsinghua University \\
yushithu@mail.tsinghua.edu.cn
\and
{\rm Shibo Zhang}\\
USSLAB, Zhejiang University \\
zhsb@zju.edu.cn
\and
{\rm Wenyuan Xu}\\
USSLAB, Zhejiang University \\
xuwenyuan@gmail.com
}
\maketitle

\footnotetext[1]{Yushi Cheng is the corresponding author.}

\begin{abstract}
    Autonomous vehicles increasingly utilize the vision-based perception module to acquire information about driving environments and detect obstacles. Correct detection and classification are important to ensure safe driving decisions. Existing works have demonstrated the feasibility of fooling the perception models such as object detectors and image classifiers with printed adversarial patches. However, most of them are indiscriminately offensive to every passing autonomous vehicle. 
    In this paper, we propose \tp, a physical adversarial patch triggered by acoustic signals. Unlike other adversarial patches, \tp remains benign under normal circumstances but can be triggered to launch a hiding, creating or altering attack by a designed distortion introduced by signal injection attacks towards cameras. To avoid the suspicion of human drivers and make the attack practical and robust in the real world, we propose a content-based camouflage method and an attack robustness enhancement method to strengthen it. Evaluations with three object detectors, YOLO V3/V5 and Faster R-CNN, and eight image classifiers demonstrate the effectiveness of \tp in both the simulation and the real world. We also discuss possible defenses at the sensor, algorithm, and system levels.
\end{abstract}

\section{Introduction}

\footnotetext[2]{Source code \& demo: https://github.com/USSLab/TPatch}

The growth of autonomous vehicles (AVs) spawns the wide deployment of automotive cameras. According to the forecast~\cite{ref}, the automotive camera market is predicted to grow at a CAGR of 12.4\% and reach a value of USD 12.2 billion by 2025. By providing precise images of surrounding environments, the automotive cameras and the subsequent perception models such as object detectors and image classifiers constitute the vision-based perception modules, which detect and classify obstacles such as cars and stop signs on the roads to help AVs make safety-critical driving decisions. As a result, correct detection and classification in the midst of a dedicated adversary are important to ensure safe driving.

However, prior work~\cite{szegedy2013intriguing,goodfellow2014explaining,brown2017adversarial,chen2018shapeshifter,athalye2018synthesizing,zhao2019seeing} has demonstrated the vulnerabilities of the vision-based perception module to carefully-crafted adversarial examples. Among the existing ones, the adversarial patch that appears in the form of localized perturbations draws much attention in the physical attacks towards AVs since it is physically realizable and robust to slight distortions. Much work~\cite{brown2017adversarial,chen2018shapeshifter,athalye2018synthesizing,zhao2019seeing} has demonstrated the effectiveness of physically-printed adversarial patches against classifiers or detectors. However, existing adversarial patches, either in the digital or physical worlds, are indiscriminately offensive to every passing AV and thus stand a chance of being detected by pilot cars equipped with recent countermeasures~\cite{chou2020sentinet,xiang2021patchguard}.

\begin{figure}[pt]
\includegraphics[width=0.95\linewidth]{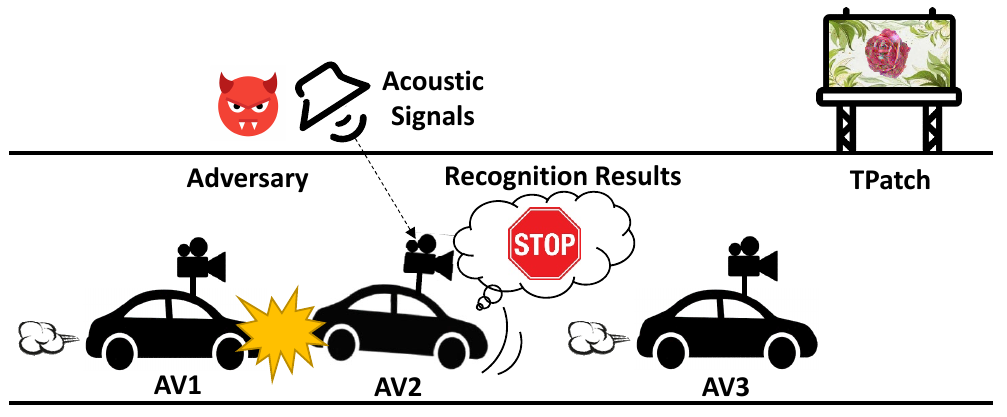}
\vspace{-0.1in}
\caption{\tp attacks. The patch is preset at the roadside, which is benign to most passing vehicles such as AV1 and AV3, but can render the targeted vehicle under acoustic injection attacks (AV2) recognize a non-existing stop sign, leading to tragic results.}
\label{fig:intro}
\vspace{-0.2in}
\end{figure}

In this paper, we ask \emph{``Can we design a physical adversarial patch that is only activated when triggered but remains benign under normal circumstances?''} Inspired by prior work~\cite{Ji2021PoltergeistAA,Sayles2021InvisiblePP} that demonstrates physical signals in addition to the natural lights may also cause distortions to images, we propose \tp, a physical adversarial patch triggered by acoustic signals. Unlike common adversarial patches, \tp remains benign under normal circumstances but can be triggered to launch a hiding, creating or altering attack toward a targeted AV by a designed distortion introduced by signal injection attacks towards cameras. 
Specifically, we consider the following attack scenario: \emph {The adversary puts a \tp at the roadside, which is benign to most passing vehicles but can render the targeted vehicle under acoustic injection attacks to recognize a non-existing stop sign, resulting in an emergency brake and thus a car collision, as shown in Fig.~\ref{fig:intro}.}

\tp is promising yet challenging. Several research challenges remain unsolved. First, how to ensure that \tp is sensitive to the image distortion caused by the selected acoustic signal while dull to other natural or unintentional distortions? Second, how to enhance the visual stealthiness of \tp and reduce the suspicion of human drivers or passengers when placed on the real road? Third, how to improve the robustness of \tp to make it more practical in the real world?

To address the aforementioned challenges, we first design a pair of triggers, where each stands for a kind of distortion, consisting of a \emph{positive trigger} that activates \tp and a \emph{negative trigger} that suppresses \tp. To find the feasible positive triggers, we estimate the image distortion caused by easy-to-implement physical signals and regularize the estimated distortion with a proposed theoretical model. Based on the selected positive and negative triggers, we design trigger-oriented loss functions for object detectors and image classifiers receptively to generate optimized \tp. Moreover, to reduce the human suspicion of the generated \tp, we propose a content-based camouflage to make the patch similar to an image or object. Finally, to improve the robustness of \tp in real-world attacks, we employ expectation over transformation and triggerable space enlargement to further enhance it.

To validate our attacks, we conduct both stimulation and real-world evaluations with three object detectors YOLO V3/V5 and Faster R-CNN, and eight image classifiers. In summary, our contributions include the points below:

\begin{itemize}
\item To the best of our knowledge, this is the first work on the physical adversarial patch that can be triggered by sensor attacks.
\item We propose a trigger-oriented optimization method to generate \tp, a content-based camouflage method to reduce the suspicion of human drivers, and an attack robustness enhancement method to further make the patch practical and robust in the real world.
\item We evaluate the effectiveness of \tp with three object detectors, YOLO V3/V5 and Faster R-CNN, and eight image classifiers in both white-box and black-box settings.
\item We realize three kinds of realistic attack implementation in the outdoor real-vehicle driving-based experiments and achieve the max success rate of 100\% and 86.4\%, w.r.t. hiding attack and creating attack, in 150 consecutive frames.
\end{itemize}

\section{Background}

In this section, we first introduce the vision-based perception module used in autonomous driving, then present adversarial patches that can fool machine learning algorithms, and finally elaborate on physical signals that may cause distortions to images.

\subsection{Vision-based Perception Module}

In autonomous driving, a vision-based perception module is critical for decision-making. The camera first converts the lights reflected from physical objects to electrical signals, processed and digitized to create digital images. Then, the perception models utilize machine learning algorithms to detect or classify objects in the images, where the perception results will be used for decision-making.

In recent years, convolutional neural networks (CNNs) based perception models have achieved high accuracy in many perception tasks. Object detectors and image classifiers are two representative types commonly used in autonomous driving. Object detectors are usually utilized to coordinate the location of critical objects, e.g., pedestrians, vehicles, etc., and classify them in a given group of classes. State-of-the-art object detectors include (1) one-stage ones represented by the YOLO series~\cite{redmon2018yolov3,yolov5} and (2) two-stage ones represented by Faster R-CNN~\cite{ren2016faster}. By contrast, image classifiers are used for more fine-grained and specialized classification tasks~\cite{apolloauto}, e.g., recognizing the states of the traffic light (red, yellow, or green) or the meanings of the traffic sign (stop, speed-limit, school, etc.)~\cite{houben2013detection}. Representative image classifiers include models of various network architectures such as VGG~\cite{simonyan2014very}, ResNet~\cite{he2016deep} and Inception~\cite{szegedy2017inception}. Both types of models provide important perception results for higher-level functions such as tracking, planning, and decision-making. In this paper, we study the vulnerabilities of both object detectors and image classifiers and their potential threats to autonomous driving.

\subsection{Adversarial Patch}

Adversarial patch~\cite{brown2017adversarial} is one type of adversarial example that appears in the form of localized perturbations. It can be a sticker attached to an existing object or a stand-alone image such as a poster. Compared to various pixel-wise perturbations directly adding to images in the digital world~\cite{szegedy2013intriguing,goodfellow2014explaining}, the adversarial patch is more likely to be practical in the real world since (1) it is physically realizable, (2) it can be placed anywhere within the field of view to cause the perception model to output a targeted class, and (3) it is image-independent and robust for rotation and scaling. Due to its practicality and robustness, it has drawn much attention recently, and a few prior works~\cite{brown2017adversarial,chen2018shapeshifter,athalye2018synthesizing,zhao2019seeing} have demonstrated the feasibility of physical adversarial patch attacks on either classifiers or detectors. However, existing adversarial patches are indiscriminately offensive to every passing vehicle and thus risk being detected by pilot cars. In this paper, we investigate the possibility of targeted adversarial attacks toward a specific victim vehicle by using a triggered patch.

\subsection{Physical Signals Affecting Imaging}

In general, the natural lights reflected from physical objects and received by the camera's image sensors are the main physical signals that affect the output images. However, other physical signals may cause distortions to the images during the imaging process, leading to misclassification of the subsequent perception models. For instance, prior work~\cite{Ji2021PoltergeistAA} has demonstrated that acoustic signals can cause controllable blur patterns on the output images by affecting the image stabilization system of the camera. Another work~\cite{Sayles2021InvisiblePP} has shown that laser diodes can inject colorful strips into the output images by exploiting the rolling shutter of the camera. These works inspire us to explore the feasibility of an adversarial patch that needs a specific image distortion caused by physical signals as a trigger to be effective. Since those physical signals, e.g., acoustic signals, lasers, etc., are commonly observed and not difficult to generate in the real world, they may be candidates for the trigger signals. In this paper, we study the blur effects caused by acoustic signals as the trigger to serve as the first example of \tp.

\section{Threat Model}

\subsection{Attack Goal}

In this paper, we consider the following attack goals:

\begin{itemize}
\item \textbf{Hiding Attacks (HA)} against object detectors that hide the detection of an existing object.
\vspace{-0.1in}
\item \textbf{Creating Attacks (CA)} against object detectors that induce the detection of a non-existing object.
\vspace{-0.1in}
\item \textbf{Altering Attacks (AA)} against image classifiers that render an object misclassified to another one.
\end{itemize}

\subsection{Adversary Capability}

To achieve the aforementioned attack goals, we assume the adversary has the following capabilities:

\textbf{Prior Knowledge of Perception Algorithms.}
We assume the adversary has prior knowledge of the object detection and image classification algorithms similar to those used in the victim AV, including their architecture, parameters, etc. Besides, the adversary can utilize the transferability of adversarial examples to achieve black-box attacks.

\textbf{Camera and Sensor Awareness.}
The adversary can acquire a camera of the same model as the one used in the victim AV, from which she can learn the information of the camera and sensors, e.g., the camera exposure time, the physical locations and parameters of the inertial sensors, etc.

\textbf{Signal Injection Capability.}
The adversary can launch acoustic injection attacks toward the inertial sensors in the victim AV by the ultrasonic transducers (1) attached to the surface of the victim AV, (2) placed at a following vehicle, or (3) placed at the roadside.

\begin{figure*}
\includegraphics[width=\linewidth]{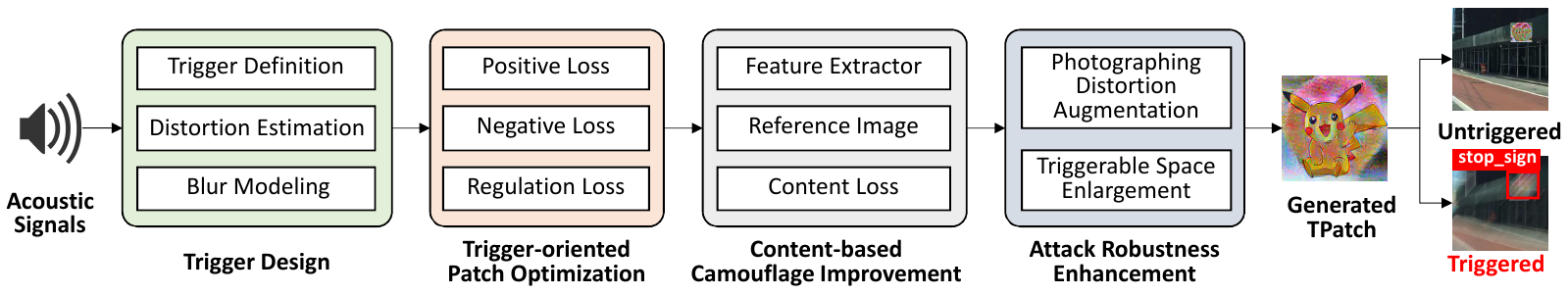}
\vspace{-0.25in}
\caption{Overview of \tp generation. Based on the selected physical trigger signal, the adversary first estimates the image distortion caused by it and designs the positive and negative triggers. Then, she trains a \tp in accord with the designed triggers, and finally improves the visual camouflage and the robustness of the patch to make it more practical in the real world. The generated \tp can then be attached to any objects to launch hiding, creating or altering attacks.}
\vspace{-0.15in}
\label{fig:overview}
\end{figure*}

\subsection{Design Requirement}

To make the attacks practical in the real world, \tp shall achieve the following design requirements:

\textbf{Conditionally Triggered.}
\tp is adversarial if and only if it is triggered by a specific physical signal. In other words, \tp shall be benign to object detectors or image classifiers when not triggered, and shall not be triggered by signals other than the designed one.

\textbf{Scene Independent.}
In the scenario of open roads, traffic participants are dynamic and hard to predict. As a result, \tp shall be adaptive to visual changes in the targeted scene, i.e., the patch shall be somewhat universal.

\textbf{Easy Implementation.}
The physical trigger signal shall be easy to generate and transmit. Thus, the patch optimization shall be trigger-oriented, i.e., we shall design an easy-to-implement signal and consider it during the patch optimization instead of finding an optimal trigger difficult to realize. In addition, \tp shall be robust to slight distortions of the trigger signal.

\section{Design}

\subsection{Overview}

In this paper, we propose \tp, a physical adversarial patch triggered by specific physical signals. Different from other adversarial patches, \tp remains benign under normal circumstances but can be triggered to launch a universal attack by a designed distortion introduced by signal injection attacks towards cameras. To craft \tp, it is important to tackle the following challenges:

\textbf{Challenge 1:} How to ensure that \tp is sensitive to the image distortion caused by the selected trigger signal while dull to other natural or unintentional distortions?

\textbf{Challenge 2:} How to enhance the visual stealthiness of \tp while keeping it functionally stealthy?

\textbf{Challenge 3:} How to improve the robustness of \tp to make it more practical in the real world?

To address these challenges, we design \tp attacks with four modules, as shown in Fig.~\ref{fig:overview}. The \textbf{Trigger Design} module first defines the positive trigger that activates \tp and the negative trigger that suppresses \tp. Then, to find the feasible positive triggers, the module estimates the image distortion caused by physical signals and regularizes the distortion with a proposed theoretical model. The \textbf{Trigger-oriented Patch Optimization} module then designs loss functions based on the selected positive and negative triggers to generate \tp capable of achieving the corresponding attack goals via gradient optimization. The \textbf{Content-based Camouflage} module further improves the visual stealthiness of \tp by applying meaningful content extracted from a reference image using a pre-trained feature extractor. The \textbf{Real-world Robustness Improvement} module improves the robustness of \tp in the real world by addressing both the deformation of the patch and the errors of the trigger signal. In the following subsections, we present the details of each module.

\subsection{Trigger Design}

\tp is designed to launch adversarial attacks when triggered by a specific physical signal while remaining benign when not triggered. To generate \tp, we first design its trigger.

\begin{figure}
\includegraphics[width=\linewidth]{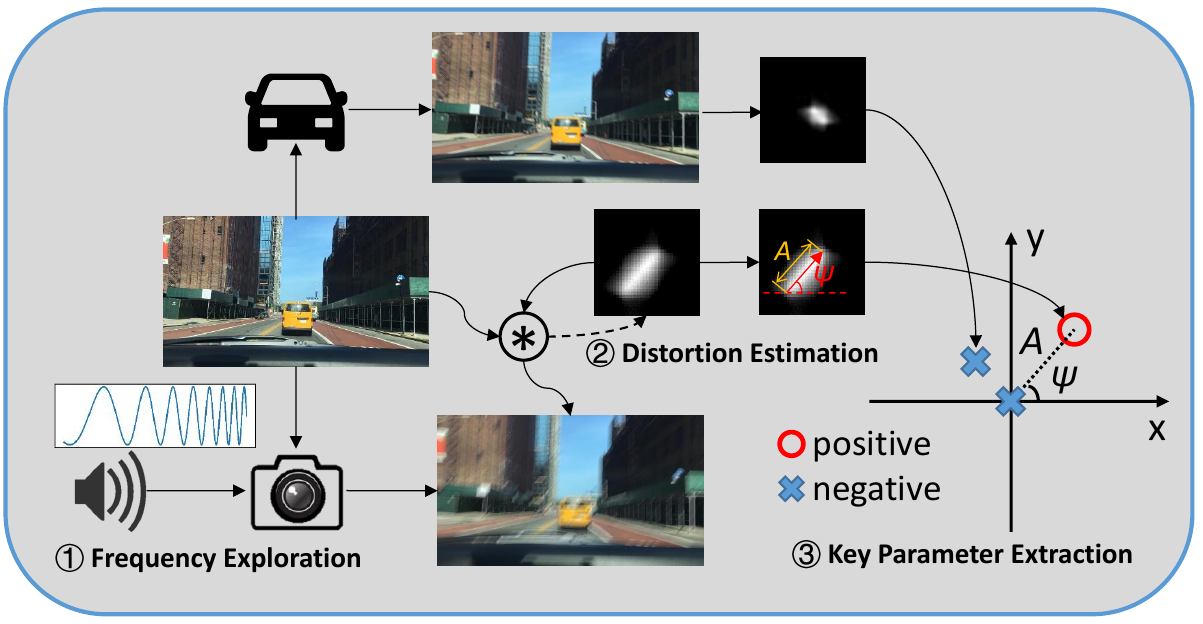}
\vspace{-0.25in}
\caption{Diagram of the trigger design. \ding{172} Explore the resonant frequency by injecting frequency-modulated acoustic signals. \ding{173} Estimate the PSF kernel with the clear image and the corresponding blurred image. \ding{174} Extract the strength and orientation of blur.}
\vspace{-0.1in}
\end{figure}

\textbf{Trigger Definition.} In this paper, we design the trigger of \tp to be a carefully crafted image blur. Specifically, we consider two opposite types of triggers: (1) positive trigger means that \tp is activated by acoustic signals and leads to adversarial detection consequences, and (2) negative trigger means that \tp is not triggered yet other factors cause undesired detection results, e.g., strong vibration caused during normal driving. 

\textbf{Signal Injection.}
We investigate which physical signal can create the expected image blur. 
Cameras rely on the outputs of motion sensors for image stabilization, which however can be manipulated by resonant acoustic signals via aliasing effects. As shown in Eq.~\ref{eq:aliasing}, the aliasing frequency $f_a$ depends on the injected signal frequency $f$ as follows: 
\vspace{-0.05in}
\begin{equation}
f_a = | f - n f_s | \leq 0.5 f_s
\label{eq:aliasing}
\end{equation}

When the injected signal frequency $f$ is an integer multiple $n$ of the sampling frequency $f_s$, the output of motion sensors, i.e., the aliased signal, is a constant bias. Thus, it can be reshaped arbitrarily through an amplification modulation~\cite{trippel2017walnut}. 

However, this method faces two major challenges in practice: (1) The phase or frequency shifting caused by the lack of synchronization may damage the attack capability in consecutive frames, as discussed in Appendix~\ref{app:challenge}. (2) The designed image blur may be difficult to implement in the real world due to the limited acoustic signal injection capability. For example, the outputs of the x/y-axis gyroscope are coupling because of their identical resonant frequencies.

To solve the first challenge, we propose a simple yet effective signal injection approach. As the resonant frequencies are a range rather than a point~\cite{trippel2017walnut}, we shift the frequency of the injected signal from the integer multiple of the sampling frequency $n f_s$ to the frequency $n f_s \pm f_e$, where $f_e$ represents the exposure frequency. In this way, the generated image blur can be resistant to phase shifting and frequency shifting, as discussed in detail in Appendix~\ref{app:approach}.

\textbf{Blur Modeling.} Then, we model the blurred image under signal injection attacks. A blurred image can be viewed as the superposition of a series of translated clean images. To simulate the blur effects, we formulate the relationship between the blurred image and the translation of image sensor, as shown in Eq.~\ref{eq:stn2}:
\vspace{-0.05in}
\begin{equation}
\begin{gathered}
\hat{u}_{i,j}^{\gamma} = \frac{1}{N} \displaystyle\sum_{t=1}^{N} Sample(i + \Delta i(t), j + \Delta j(t); v^{\gamma}) \\
\Delta i(t) = A \cos \psi \sin (2 \pi \frac{t}{N} + \phi_i) \\
\Delta j(t) = A \sin \psi \sin (2 \pi \frac{t}{N} + \phi_j)
\end{gathered}
\label{eq:stn2}
\end{equation}
where $\hat{u}$ and $v$ denote the blurred and clean image, respectively. $i$ and $j$ represent the geometric position of a pixel. $\Delta i(t)$ and $\Delta j(t)$ represent the translation of the camera, which is in a form of sine wave with the similar phases $\phi_i, \phi_j$ but different amplitudes $A \cos \psi, A \sin \psi$. $Sample(\cdot;\cdot)$ denotes the sampling function by interpolation, e.g., bilinear. $N$ represents the fineness of discretion. $\gamma =2.2$ is the parameter of gamma correction~\cite{anderson1996proposal}, which converts the RGB values into the luminosity that is linear with exposure duration. The strength $A$ and the orientation $\psi$ are the key parameters of blur modeling. The slight phase difference $\Delta \phi = |\phi_i - \phi_j|$ has limited effects on \tp as shown in Fig.~\ref{fig:phase-diff} in Appendix~\ref{app:phase}.

\textbf{Parameter Estimation.}
In the prior work, the attacker first optimizes the desired blur by simulation and then implements it in the real world. However, not all the optimized blur can be realized due to the coupling of the x/y-axis gyroscope. 
To ease the burden of the physical attack and make it more practical in the real world, we propose an estimation-based method to obtain the possible image blur set by injecting acoustic signals easy to implement. In general, the image blur caused by motions (introduced by acoustic signals) can be described with a convolutional kernel named Point Spread Function (PSF)~\cite{fergus2006removing}, as shown in Eq.~\ref{eq:psf}:
\vspace{-0.03in}
\begin{equation}
    \hat{u}_{i,j}^{\gamma} = \sum \limits_{a=-k}^{k} \sum \limits_{b=-k}^{k} v_{i-a,j-b}^{\gamma} h_{a,b}
    \label{eq:psf}
\end{equation}
where $h$ denotes the PSF, whose kernel size is $(2k+1) \times (2k+1)$ and $k$ is an integer. $\hat{u}$ is the convolutional result of the PSF and $v$ is the clean image. To estimate the kernel, we formulate it as an optimization problem, as shown in Eq.~\ref{eq:opt}:
\vspace{-0.03in}
\begin{equation}
    \min_h \lVert u - \hat{u} \rVert_2 = \min_h \lVert u - (v^{\gamma} \ast h)^{1/\gamma} \rVert_2
    \label{eq:opt}
\end{equation}
where $u$ denotes the blurred image captured from a physical signal injection attack (the ground truth), and symbol $*$ denotes the convolutional operator. We use the Mean-Square-Error (MSE) between the convolutional result $\hat{u}$ and the real-captured image $u$ as the loss function and solve the optimization problem via gradient descent. In this way, we obtain the estimated PSF $h$ and calculate its strength $A$ and orientation $\psi$ as the parameters of a possible image blur.

\subsection{Trigger-oriented Patch Optimization}
 
With the designed positive and negative triggers, we then optimize \tp based on them. Note that in contrast to common adversarial patches that have a straightforward goal to fool image classifiers, object detectors, or other victim systems, \tp is benign until it is triggered. As a result, we consider several losses when design the loss functions: (1) a positive loss ${L}_{pos}$ that makes the patch achieve the adversarial effects with a positive trigger, (2) a negative loss ${L}_{neg}$ that makes the patch dull to any negative triggers, (3) a regulation loss ${L}_{TV}$ that reduces the risk of over-fitting and enhances the capability of migrating the attack from the digital world to the physical world, and (4) a content loss ${L}_{content}$ that realizes a meaningful camouflage on the generated patch (details can be found in Sec.~\ref{subsec:content}). The integral loss function is shown in Eq.~\ref{eq:overall-loss}:
\vspace{-0.03in}
\begin{equation}
    \mathop{\arg \min}\limits_{p} \mathbb{E}_{x \sim X,l \sim L,t \sim T}\left[ {L}_{pos} + \lambda {L}_{neg} \right] + \alpha {L}_{TV} + \beta {L}_{content}
    \label{eq:overall-loss}
\end{equation}
where $p$ is the \tp. $x,l,t$ represent the input image, the location where the patch is applied, and the random transformation respectively. Then, $X,L,T$ denote the corresponding distribution. $\mathbb{E}$ denotes the expectation of the loss. Hyperparameters, e.g., $\lambda$, $\alpha$, $\beta$, are used to balance different loss components.

The positive and negative losses, i.e., $L_{pos}$ and $L_{neg}$, are different across object detectors and image classifiers. The classifier outputs a single vector that represents the probability of each class while the detector predicts the objectiveness scores, the classification scores, and the object's location. For classifiers, the aggregation of features enables the patch to manipulate all the outputs. However, for detectors that have multiple predictive cells to recognize multiple objects, the patch's location mainly determines the affecting area due to the locality of the patch. As a result, we shall align the target object's location with the adversarial patch to achieve an effective attack. What's more, the losses vary as the attack goal changes. Therefore, we discuss the detailed design of the loss function concerning each attack type.

For HA, the adversarial case is to render the target object with the \tp undetected, while the benign one is to keep the target object correctly recognized. The loss functions of HA are formulated as Eq.~\ref{eq:loss-ha}.

\begin{equation}
\begin{aligned}
{L}_{pos} &= -\log (1 - \max (p_{obj} \cdot p_{t})) \\
{L}_{neg} &= -\log \left(p_{obj} \cdot p_{t}\right) + \zeta L_{reg}
\end{aligned}
\label{eq:loss-ha}
\end{equation}
where $p_t$ and $p_{obj}$ denote the classification scores of target class $t$ and the objectiveness scores respectively. $L_{reg}$ is the regression loss to guide the detected box. $\zeta$ is the hyperparameter to balance the recognition loss and regression loss.

For CA, the adversarial case is to render the \tp recognized as the target object, while the benign one is to keep the \tp undetected. The loss functions of CA are formulated as Eq.~\ref{eq:loss-ca}.

\begin{equation}
\begin{aligned}
{L}_{pos} &= -\log \left(p_{obj} \cdot p_{t}\right) + \zeta L_{reg} \\
{L}_{neg} &= -\log (1 - \max (p_{obj} \cdot p_{t}))
\end{aligned}
\label{eq:loss-ca}
\end{equation}

For AA, the adversarial case is to render the images with the \tp recognized as the target class, while the benign one is to keep the original recognition results. The loss functions of AA are formulated as Eq.~\ref{eq:loss-aa}.

\begin{equation}
\begin{aligned}
{L}_{pos} &= -\log \left({p}_{t}\right) \\
{L}_{neg} &= -\log \left(1-{p}_{t}\right)
\end{aligned}
\label{eq:loss-aa}
\end{equation}
where $p_t$ is the probability of target class $t$ predicted by the classifier.

The total variation loss is used for regularizing the \tp. As shown in Eq.~\ref{eq:loss-tv}, the TV loss aims to reduce the color changes between the adjacent pixels, which eliminates the overfitting of \tp in the digital simulation as well as improves the image quality. 

\begin{equation}
{L}_{TV}  = \sum_{i, j} \sqrt{\left|x_{i+1, j}-x_{i, j}\right|^{2}+\left|x_{i, j+1}-x_{i, j}\right|^{2}}
\label{eq:loss-tv}
\end{equation}

\subsection{Content-based Camouflage Improvement}
\label{subsec:content}

\begin{figure}
\includegraphics[width=0.95\linewidth]{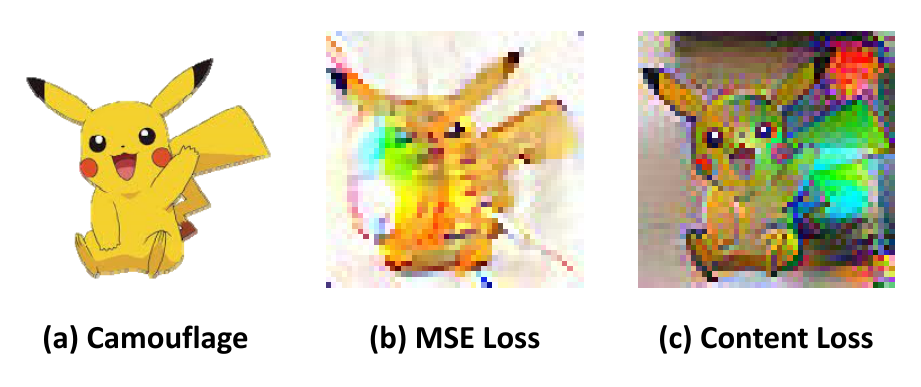}
\vspace{-0.15in}
\caption{Comparison between MSE loss and content loss. (a) shows the target camouflage of the patch. (b) and (c) show the generated patch by MSE loss and content loss respectively.}
\vspace{-0.1in}
\label{fig:content}
\end{figure}

Common adversarial patches are usually generated without constraints on the $L_2 / L_{\infty}$ for the sake of universality, and thus are inevitably perceptible to human eyes. Although prior work has demonstrated the feasibility of changing the shape of the patch arbitrarily, it remains abnormal since the pattern of the patch is meaningless. To address it, we propose the content-based camouflage improvement to make the patch more meaningful, e.g., to be like a cartoon character or an art painting.

A vanilla approach is to add a loss function that constrains the patch with a euclidean distance between the target image and the patch. However, this vanilla approach leads to a tricky problem regarding the choice of the target image, i.e., the punishment of the loss varies dramatically with different target images. Besides, the vanilla loss function is color-based, which means the outline of the camouflage is prone to diminish, rendering the camouflage ambiguous and over-smooth.

To improve the camouflage, we utilize a content loss proposed by style transfer works~\cite{gatys2016image,johnson2016perceptual} to regularize the patch based on high-level features extracted by a pre-trained CNN. The content loss can be formulated as Eq.~\ref{eq:content-loss}. We use the $j$-th layer $\phi_{j}$ of the CNN pre-trained on ImageNet to respectively extract the high-level feature map of target image $\hat{x}$ and patch $x$ with the same shape $C_{j} \times H_{j} \times W_{j}$. The loss is calculated with the euclidean distance in the feature space rather than the pixel space. It encourages the patch to learn the content and spatial structure of the target image rather than the details, e.g., color, texture, and exact shape. A comparison of \tp with the proposed content loss and the conventional MSE loss is shown in Fig.~\ref{fig:content}, from which we can see that the content loss helps \tp to be more natural as a cartoon character.

\begin{equation}
L_{content} = \frac{1}{C_{j} H_{j} W_{j}}\left\|\phi_{j}(\hat{x})-\phi_{j}(x)\right\|_{2}^{2} \label{eq:content-loss}
\end{equation}

\subsection{Attack Robustness Enhancement}
\label{subsec:robust}

The challenges of robust \tp attacks mainly lie in two aspects: (1) the captured patch can be different from the digital one due to the additional process of printing and filming, and (2) the injected trigger can be coarse and noisy due to the limited physical attack capability and the estimation errors. The former challenge is shared with other adversarial patches. Factors commonly considered affecting the filming include lighting conditions, shooting distances and angles. The latter is unique for \tp since it requires a certain trigger to be adversarial. Inevitably, trigger errors can be introduced in the pipeline of physical signal injection, trigger estimation, and trigger formulation. The difference between the estimated trigger and the ground truth probably causes the failure of activating the \tp. To tackle these challenges, we utilize the expectation over transformation for the former and the triggerable region enlargement for the latter.

\textbf{Expectation over Transformation.}
To address the patch distortion caused by photographing, we use the Expectation over Transformation (EoT) method~\cite{athalye2018synthesizing}, which augments the training of \tp with random transformation to overcome the various situations in the real world. We augment the patch with three dimensions, i.e., resize, rotation, and color shift (including brightness, contrast, saturation, and hue). We perform three different transformations simultaneously with a uniform distribution to randomize the degree of each transformation.

\begin{figure}
\centering
\includegraphics[width=.9\linewidth]{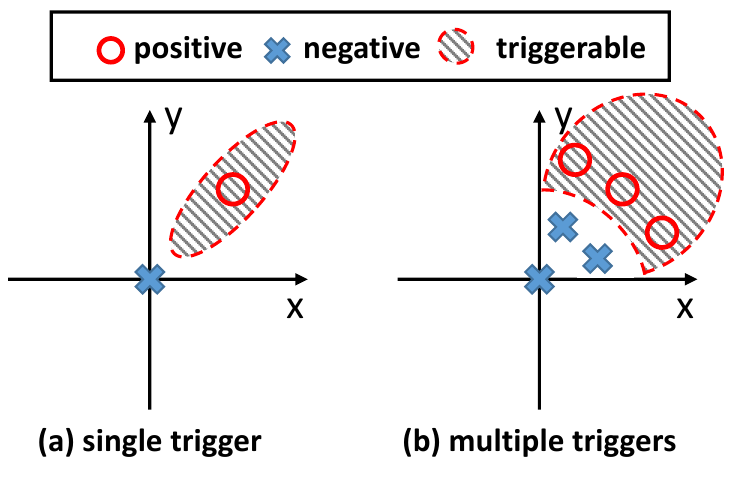}
\vspace{-0.1in}
\caption{Illustration of the relationship between the triggerable region and the designed triggers.}
\label{fig:trigger}
\end{figure}

\textbf{Triggerable Region Enlargement.}
Since an adversarial subspace usually includes more than one point, 
triggers close to the designed positive/negative trigger can be positive/negative as well, according to our observation. 
Inspired by this phenomenon, we propose to enlarge the triggerable region by employing more positive and negative triggers,  as illustrated in Fig.~\ref{fig:trigger}. Such an operation helps avoid the mild blur existing in normal driving from activating \tp and improves the robustness of \tp with more possible directions and strengths of the blur.

\section{Evaluation}

In this section, we evaluate \tp against both object detectors and image classifiers. We consider two sets of evaluations in this paper: (1) simulation evaluation, where \tp is attached to digital images directly, and (2) real-world evaluation, where \tp is physically printed to conduct attacks, and the experiments are conducted with the outdoor real-vehicle driving-based setup. We use the attack success rate (ASR) as the metric to evaluate the simulation experiments and use the highest attack success rate in \textit{n} consecutive frames $f_{succ}^{max(n)}$~\cite{zhao2019seeing,shen2022sok} to evaluate the real-world experiments.

In summary, we highlight the key results of our attacks as follows:

\begin{itemize}
    \item In the simulation evaluation, \tp can achieve overall ASRs of 88.4\% and 91.9\% for HA and CA against three object detectors YOLO V3/V5 and Faster R-CNN, and achieve an overall ASR of 85.7\% for AA against eight image classifiers.
    \item In the real-world evaluation, \tp can achieve overall $f_{succ}^{max(150)}$ of 100.0\% and 86.4\% for HA and CA against three object detectors YOLO V3/V5 and Faster R-CNN.
    \item We conduct attacks in three attack scenarios, i.e., in-car, car-to-car, and roadside, with an autonomous driving vehicle on closed roads. The farthest feasible attack range reaches about 6 meters.
    \item We extensively investigate the transferability and effectiveness of \tp under various driving scenarios in the real world. 
\end{itemize}

\subsection{Simulation Evaluation}

In the simulation evaluation, we investigate three types of attacks, i.e., hiding attack (HA), creating attack (CA), and altering attack (AA), in the public datasets extensively. We first evaluate multiple classes of interest for each attack under a consistent setting. Then we investigate the impact of critical factors such as patch sizes and trigger parameters on the effectiveness of \tp. Finally, we study the transferability of \tp with two black-box attacks, i.e., single model attack and ensemble model attack.

\subsubsection{Experimental Setup}

\textbf{Object Detectors.} We evaluate \tp using three popular object detectors, including both one-stage ones YOLO V3/V5 and a two-stage one Faster R-CNN. The backbones of the pre-trained models Faster R-CNN, YOLO V3, and YOLO V5 are ResNet-50, Darknet-53, and CSPDarknet, respectively. All the three models are trained on the MS Common Objects in Context (COCO) dataset~\cite{coco} for detection. 

\textbf{Image Classifiers.} We evaluate \tp on eight widely-used CNNs, i.e., VGG-13/16/19, ResNet-50/101/152, Inception v3, and MobileNet v2, which cover models of different depths and architectures. All of them are trained on the large vision database ImageNet~\cite{deng2009imagenet} for classification. 

\textbf{Datasets.} For object detectors, we utilize two popular autonomous driving datasets KITTI~\cite{geiger2012we} and BDD100K~\cite{yu2020bdd100k} that consist of images captured from real driving scenarios for evaluation. For the image classifier, we utilize the ImageNet validation set for evaluation. We use 2,000 and 10,000 images that are unseen in the training of the \tp for testing detectors and classifiers, respectively.

\textbf{Classes of Interest.} We select 8 classes for object detectors and 20 classes for image classifiers as the classes of interest. The classification or detection results of these classes are security-related in the scenario of autonomous driving.

\textbf{Metrics.}
We use the attack success rate (ASR) as the metric in the simulation evaluation, which is the ratio of the number of successful attacks against an object detector or image classifier over the total number of conducted attacks. We define a \tp attack to be successful when it achieves the targeted adversarial effect under triggering signals while it does not modify the detection results without triggering. With this definition, the metric can be formulated as Eq.~\ref{eq:metric}:
\begin{equation}
    ASR = \frac{1}{N} \sum_{i=1}^{N} \mathbb{I}_{F\left(x, {tr}_p \right) = y_a \& F\left(x, {tr}_n \right) = y_b} (x)
    \label{eq:metric}
\end{equation}
where $\mathbb{I}(x)$ is the indicator function. $F$ means the recognition function of the target model. $x$ means the image embedded with \tp. $tr_p$ and $tr_n$ are the positive trigger and negative trigger respectively. $y_a$ and $y_b$ denote the adversarial and benign recognition results respectively.


More concretely, the predicted class is always the one that has the highest confidence, and the prediction of \tp is decided by the Intersection-over-Union (IoU) between any bounding box with the bounding box of \tp, where we use the threshold 0.5 to generate a binary result. In accord with common sense, we refer to the object attached with the \tp as the targeted box rather than the \tp itself when evaluating HA.

\subsubsection{Overall Performance}

\begin{figure}[pt]
    \includegraphics[width=.95\linewidth]{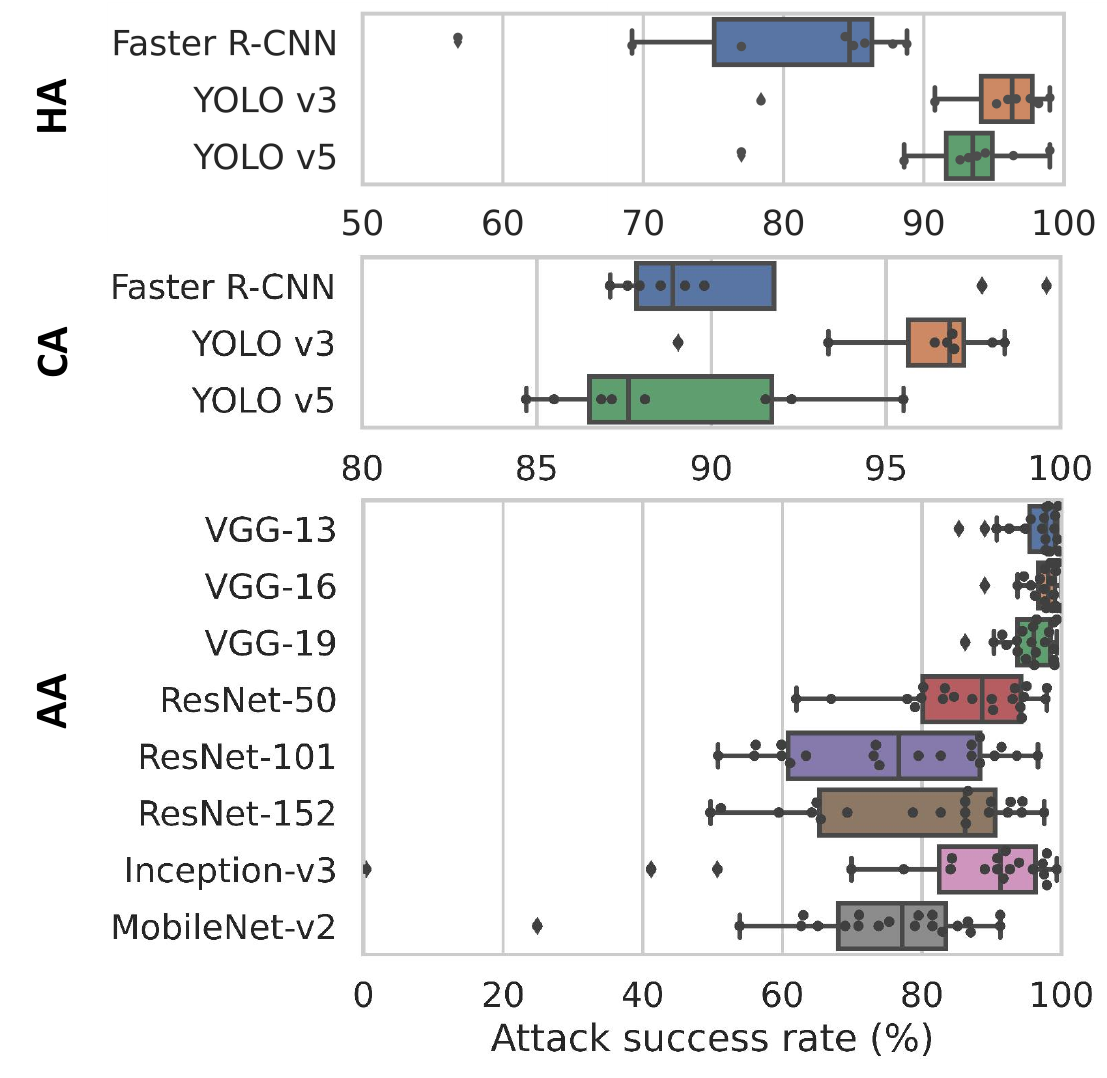}
    \vspace{-0.1in}
    \caption{Overall effects of \tp with regard to HA (top), CA (middle) and AA (bottom). The box plots show the min/max and quarterlies, and the dots represent different classes of interest.}
    \vspace{-0.2in}
    \label{fig:overall}
\end{figure}

In this subsection, we evaluate the effectiveness of \tp attacks on different detectors and classifiers, respectively. During the evaluation, we used a pair of positive and negative triggers. The strength of the positive and negative triggers are 8 pixels and 0 pixels, respectively, and the orientation of the positive trigger is 45\degree. The patch size is set to $48 \times 48$ pixels in the overall evaluation. 

For every single test of CA and AA, we randomly place the \tp on a single image and then use a pair of the designed positive and negative triggers to generate two images that stand for the adversarial and benign cases for recognition. According to the metric formulated in~\ref{eq:metric}, the overall success rate is calculated. Unlike the other attacks, the \tp of HA is attached to the target object rather than standalone. According to its annotations, the target objects are cropped from the MS COCO dataset. The appropriate size bounds filter them to ensure the occupied proportion, round 20\% of the object's area, of the patch.

The results of overall performance evaluation are shown in Fig.~\ref{fig:overall}. For HA, the overall success rates are 79.4\%, 94.0\%, and 91.9\%, w.r.t. Faster R-CNN, YOLO V3 and YOLO V5. For CA, the overall success rates are 91.0\%, 95.7\%, and 89.0\%, w.r.t. Faster R-CNN, YOLO V3 and YOLO V5. For AA, the VGG series are the easiest to attack (over 95\%), followed by ResNet and Inception (over 75\%), and the worst is MobileNet (73.8\%). We find that the results have notable differences across classes, shared with different classifiers and detectors. For example, in the experiments of Inception v3, the most vulnerable class is \textit{traffic light}, which exceeds the 99\% success rate. In contrast, the least vulnerable class, \textit{trolleybus}, only achieves a 0.04\% success rate, which means it hardly has adversarial effects. In addition, it is observed that the worst classes are shared with different models, e.g., \textit{trolleybus}, \textit{pickup}, etc. The same phenomenon is observed in the HA and CA experiments on detectors. For instance, \textit{car} is the easiest to achieve CA, and \textit{person} is the most difficult to achieve HA in all the detectors.

\subsubsection{Impact of Patch Details}

In this subsection, we investigate several possible factors that may influence the adversarial effects of \tp, including the size of the patch and the strength and orientation of the trigger.

For simplicity, we use several representative models for patch detail analysis. As for the image classifier, we choose the ResNet-50, whose overall performance lies in the medium position among the eight models, and the \textit{traffic light} as the attack target of AA. As for the object detector, we choose the YOLO V5, which is one of the latest models, and the \textit{stop sign} as the attack target of CA, which keeps consistency with the real-world evaluation.

\begin{figure}
    \includegraphics[width=.95\linewidth]{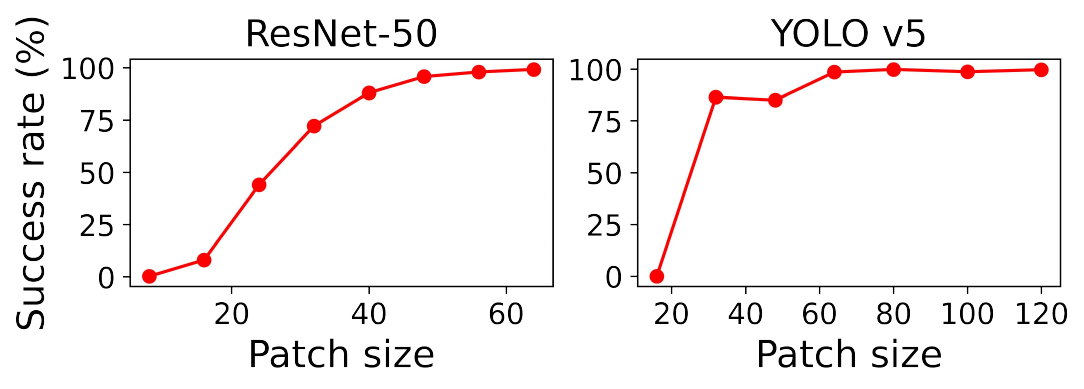}
    \vspace{-0.1in}
    \caption{Attack success rates under different patch sizes. The shape of patch is square, and the patch size is denoted by the side length of the square.}
    \vspace{-0.2in}
    \label{fig:size}
\end{figure}

\textbf{Impact of Patch Size.}
The patch size represents the capability of an attacker to manipulate the input image. In the autonomous driving scenario, the small patch size indicates the far distance between the camera and the patch, and the large patch size represents that the camera is near the patch. As shown in Fig. \ref{fig:size}, too tiny patches hardly achieve the success of the attack. The success rate curve is relatively smooth in the classifier and achieves the success rate of nearly 100\% when the patch size is $64 \times 64$ pixels, while two leaps of success rate in the detector can be observed at $32 \times 32$ pixels and $64 \times 64$ pixels. The reason is related to the three scales in YOLO V5. As the patch size arises, it is more likely to be wrongly predicted by the detector.

\begin{figure}
    \includegraphics[width=\linewidth]{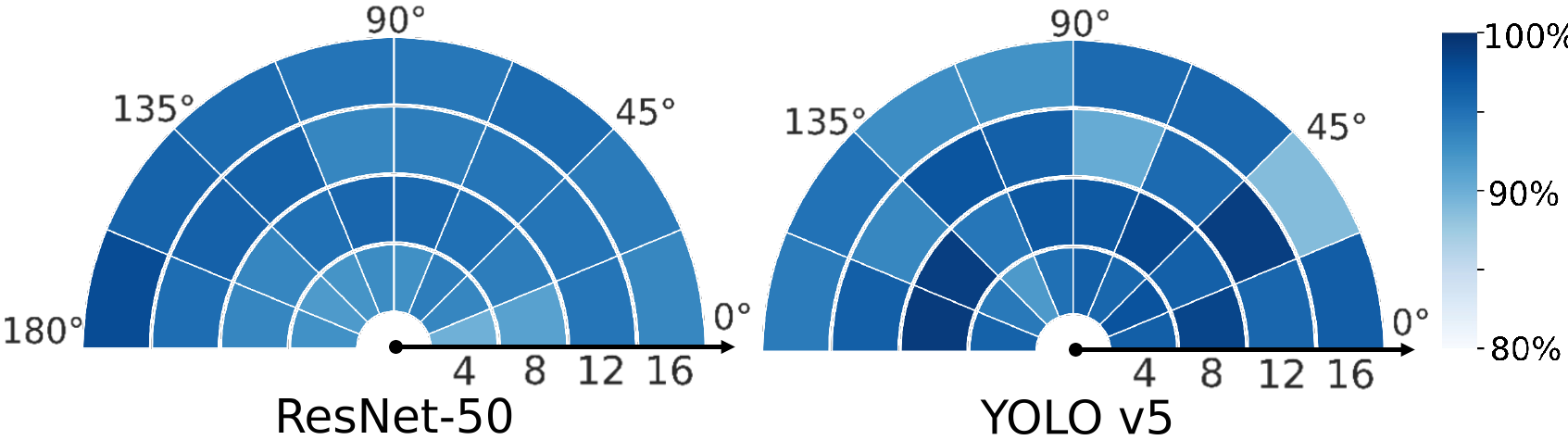}
    \vspace{-0.15in}
    \caption{Attack success rates with different triggers. The strength and orientation of blur trigger is illustrated by polar coordinates.}
    \label{fig:tri}
    \vspace{-0.1in}
\end{figure}

\textbf{Impact of Triggers.}
To investigate whether different image blurs would influence the adversarial effect of \tp, which is crucial for our trigger-oriented optimization, we set different positive triggers whose strength and orientation are diverse. Due to the symmetrical effects of linear blur, we design the range of orientation from 0 to 180 degrees, and we set the range of strength from 4 to 16 pixels. Meanwhile, we keep other parameters as default in the overall experiment. As shown in Fig. \ref{fig:tri}, the \tp achieves more than 88\% success rate in every blur pattern with different orientations and strengths. Thus, it is demonstrated that the effectiveness of \tp does not rely on the specific blur.

\subsubsection{Stealthiness Study}

\begin{figure}[pt]
\centering
\includegraphics[width=.95\linewidth]{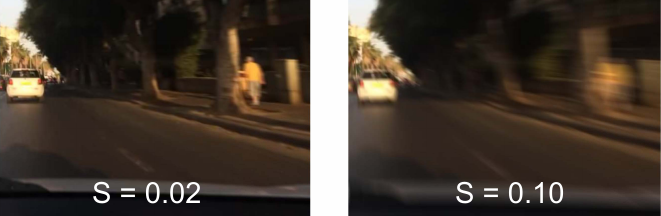}
\caption{Images with different degrees of radial blur. The degrees of radial blur equal 0.02 (left) and 0.1 (right).}
\label{fig:radial}
\vspace{-0.1in}
\end{figure}

\begin{table}[pt]
\centering
\caption{Attack success rates under different triggers.}
\label{tab:neg-tr}
\scalebox{0.85}{
\begin{tabular}{|c|c|c|c|c|c|c|}
    \hline
    \diagbox{$A$}{$\psi$} 
    & 15\degree & 45\degree & 75\degree & 105\degree & 135\degree & 165\degree
    \\ \hline
    4.5 & 0\%   & 0\%       & 0\%       & 0\%        & 0\%        & 0\%
    \\ \hline
    9   & 0\%   & 5\%       & 10\%      & 0\%        & 0\%        & 0\%
    \\ \hline
    18  & \cellcolor{gray}54\% & \cellcolor{gray}97\% & \cellcolor{gray}87\% 
    & 0\% & 0\% & 0\%
    \\ \hline
\end{tabular}}
\vspace{-0.1in}
\end{table}

The captured images are not always clear because of the car movements and vibrations. To investigate the functional stealthiness of \tp, we simulate the possible blur effects via two methods, i.e., (1) radial blur for car movements and (2) other triggers for car vibrations.

\textbf{Car Movements.} We use radial blur to simulate the blur effects caused by car movements. To measure the boundary of the radial blur, we drive at a speed of 40-60 km/h in the low light scenario meanwhile sampling real-world blur effects. We compare the captured images with the simulated images to estimate the degree of radial blur $S \approx 0.02$. Then we test the \tp for CA attacks on 500 images with different degrees of radial blur. The experiment results show that the mis-triggering rate is 0\% when $S = 0.02$ and achieves 5.2\% when $S = 0.1$ (corresponding to a car speed around 200-300 km/h), as illustrated in Fig.~\ref{fig:radial}. Thus, the radial image blur caused by a moving car at a reasonable speed may not trigger the \tp. 

\textbf{Car Vibrations.} Since the car vibrations are various in the real world, we simulate the effects by investigating the ASRs with triggers of various strengths $A$ and orientations $\psi$. Tab.~\ref{tab:neg-tr} outlines the ASRs of \tp under different triggers. The shadowed cells represent the designed positive triggers, while the rest are negative triggers. We find the following observations. In most cases, \tp cannot be mis-triggered, and the exceptions are a few cases that have similar strengths $A$ and orientations $\psi$ as the positive ones. Nevertheless, the average mis-triggering rate is only 1\%.

\subsubsection{Transferability Study}

\begin{table*}[pt]
\caption{Transferability across different detectors.}
\label{tab:trans1}
\vspace{-0.1in}
\begin{center}
\setlength{\tabcolsep}{6mm}{
\scalebox{0.80}{
\begin{tabular}{c|c|c|c|c|c|c}
    \hline
    Attack Type  
    & \multicolumn{3}{c|}{Hiding Attack} 
    & \multicolumn{3}{c}{Creating Attack} 
    \\ \hline
    Detector 
    & Faster R-CNN  & YOLO V3 & YOLO V5   
    & Faster R-CNN  & YOLO V3 & YOLO V5 
    \\ \hline \hline
    Faster R-CNN 
    & 85.8\%        & 43.7\%  & 42.8\%  
    & 97.9\%        & 31.2\%  & 46.3\%    
    \\ \hline
    YOLO V3 
    & 63.8\%        & 98.9\%  & 86.2\%  
    & 34.9\%        & 99.1\%  & 87.1\%    
    \\ \hline
    YOLO V5 
    & 70.1\%        & 78.6\%  & 93.8\%  
    & 61.5\%        & 93.8\%  & 99.8\%    
    \\ \hline
\end{tabular}}}
\end{center}
\vspace{-0.3in}
\end{table*}

\begin{table*}[pt]
\caption{Transferability across different classifiers.}
\label{tab:trans2}
\vspace{-0.1in}
\begin{center}
\setlength{\tabcolsep}{6mm}{
\scalebox{0.80}{
\begin{tabular}{c|c|c|c|c|c|c|c|c}
    \hline
    Classifier & VGG13  & VGG16  & VGG19  & Res50  & Res101  & Res152  & Incv3  & Mobv2  \\ \hline \hline
    VGG-ens    & 94.0\% & 96.0\% & 96.2\% & 42.7\% & 32.0\%  & 49.4\%  & 34.3\% & 32.8\% \\ \hline
    Res-ens    & 26.1\% & 51.8\% & 51.9\% & 91.8\% & 90.9\%  & 91.4\%  & 52.4\% & 45.6\% \\ \hline
\end{tabular}}}
\end{center}
\vspace{-0.2in}
\end{table*}

When the adversary has very limited prior knowledge of the DNN models used in commercial autonomous cars, the gradient-based optimization approach can not be directly applied to those black-box models. However, the attacker is still possible evade the target model via the transferability of adversarial examples among similar DNN models. To evaluate the transferability of \tp, we conduct a single model attack on detectors and an ensemble model attack on classifiers. During the evaluation, the design details of \tp, e.g., target class, patch size, trigger setting, etc., is all fixed except for the recognition with different models. 

Tab.~\ref{tab:trans1} list the results of transfer attacks on different detectors. Due to the similar model architecture, the success rates of transfer attacks between YOLO V3 and YOLO V5 are relatively high, 86\% (average), than the one between the one-stage detectors (YOLO series) and the two-stage detectors (Faster R-CNN), 49\% (average), in all the HA and CA experiments. As we use the same metric as formulated in Eq.~\ref{eq:metric}, the suppression effect from negative triggers is confirmed in the black-box models in addition to the adversarial effect from positive triggers. 

We investigate the ensemble model attack on different classifiers. Given that the attacker has a series of white-box models with the same architecture while attacking a black-box model, we build two ensemble models, i.e., VGG-ens (VGG13+VGG16+VGG19) and Res-ens (Res50+Res101+Res152), as the white-box model to optimize the \tp to launch transfer attacks on the other five models. Tab.~\ref{tab:trans2} list the detailed success rates of ensemble transfer attacks. The VGG-ens and Res-ens attack achieve a 38\% and 46\% average success rate on five black-box models. Since the number of classes in ImageNet, 1000, is much more than the number of classes in MS COCO, 80, the overlap of decision space between two different models is smaller theoretically; thus, the targeted transfer attack would be more challenging.

\subsection{Real-world Evaluation}

\begin{table*}[pt]
\caption{Overall performance of hiding attacks and creating attacks in real-world setups.}
\label{rw:overall-ca}
\vspace{-0.05in}
\begin{center}
\centering
\setlength{\tabcolsep}{5.5mm}{
\scalebox{0.80}{
\begin{tabular}{cc|ccc|ccc}
    \hline
    \multicolumn{2}{c|}{Attack Type} & \multicolumn{3}{c|}{Hiding Attack} & \multicolumn{3}{c}{Creating Attack} \\ \hline
    \multicolumn{1}{c|}{Source}      & Target      & \multicolumn{1}{c|}{$f_{succ}^{max(50)}$} & \multicolumn{1}{c|}{$f_{succ}^{max(100)}$} & $f_{succ}^{max(150)}$ & \multicolumn{1}{c|}{$f_{succ}^{max(50)}$} & \multicolumn{1}{c|}{$f_{succ}^{max(100)}$} & $f_{succ}^{max(150)}$ \\ \hline 
    \hline
    \multicolumn{1}{c|}{Faster R-CNN} & Faster R-CNN 
    & \multicolumn{1}{c|}{100.0\%} & \multicolumn{1}{c|}{100.0\%} & 100.0\% 
    & \multicolumn{1}{c|}{100.0\%} & \multicolumn{1}{c|}{90.0\%} & 64.0\%
    \\ \hline
    \multicolumn{1}{c|}{Faster R-CNN} & YOLO V3     
    & \multicolumn{1}{c|}{100.0\%}  & \multicolumn{1}{c|}{100.0\%} & 98.0\% 
    & \multicolumn{1}{c|}{96.0\%}  & \multicolumn{1}{c|}{65.0\%} & 46.7\% 
    \\  \hline
    \multicolumn{1}{c|}{Faster R-CNN} & YOLO V5 
    & \multicolumn{1}{c|}{100.0\%} & \multicolumn{1}{c|}{98.0\%} & 96.7\% 
    & \multicolumn{1}{c|}{30.0\%} & \multicolumn{1}{c|}{27.0\%} & 20.0\% 
    \\  \hline
    \multicolumn{1}{c|}{YOLO V3}   & Faster R-CNN 
    & \multicolumn{1}{c|}{100.0\%} & \multicolumn{1}{c|}{100.0\%} & 100.0\% 
    & \multicolumn{1}{c|}{100.0\%} & \multicolumn{1}{c|}{95.0\%} & 92.7\% 
    \\  \hline
    \multicolumn{1}{c|}{YOLO V3}   & YOLO V3 
    & \multicolumn{1}{c|}{100.0\%} & \multicolumn{1}{c|}{100.0\%} & 100.0\% 
    & \multicolumn{1}{c|}{100.0\%} & \multicolumn{1}{c|}{100.0\%} & 100.0\% 
    \\  \hline
    \multicolumn{1}{c|}{YOLO V3} & YOLO V5 
    & \multicolumn{1}{c|}{100.0\%} & \multicolumn{1}{c|}{98.0\%} & 96.0\% 
    & \multicolumn{1}{c|}{100.0\%} & \multicolumn{1}{c|}{79.0\%} & 77.3\% 
    \\  \hline
    \multicolumn{1}{c|}{YOLO V5}     & Faster R-CNN 
    & \multicolumn{1}{c|}{100.0\%} & \multicolumn{1}{c|}{100.0\%} & 100.0\% 
    & \multicolumn{1}{c|}{96.0\%} & \multicolumn{1}{c|}{82.0\%} & 71.3\% 
    \\  \hline
    \multicolumn{1}{c|}{YOLO V5}     & YOLO V3     
    & \multicolumn{1}{c|}{100.0\%} & \multicolumn{1}{c|}{100.0\%} & 99.3\% 
    & \multicolumn{1}{c|}{90.0\%} & \multicolumn{1}{c|}{83.0\%} & 77.3\%  
    \\  \hline
    \multicolumn{1}{c|}{YOLO V5}     & YOLO V5     
    & \multicolumn{1}{c|}{100.0\%} & \multicolumn{1}{c|}{100.0\%} & 100.0\% 
    & \multicolumn{1}{c|}{100.0\%} & \multicolumn{1}{c|}{98.0\%} & 95.3\%
    \\  \hline
\end{tabular}}}
\end{center}
\vspace{-0.3in}
\end{table*}

In the real-world evaluation, we focus on these two kinds of attacks, i.e., HA and CA against object detectors under both white-box and black-box settings. We implement attacks with three scenarios, i.e., in-car, car-to-car, and roadside. Furthermore, we extensively investigate the influence of various factors on the attack performance, including attack distances, victim cameras, vehicle speeds, road conditions, and lighting conditions.

\subsubsection{Experimental Setup}

\begin{figure}[pt]
    \includegraphics[width=\linewidth]{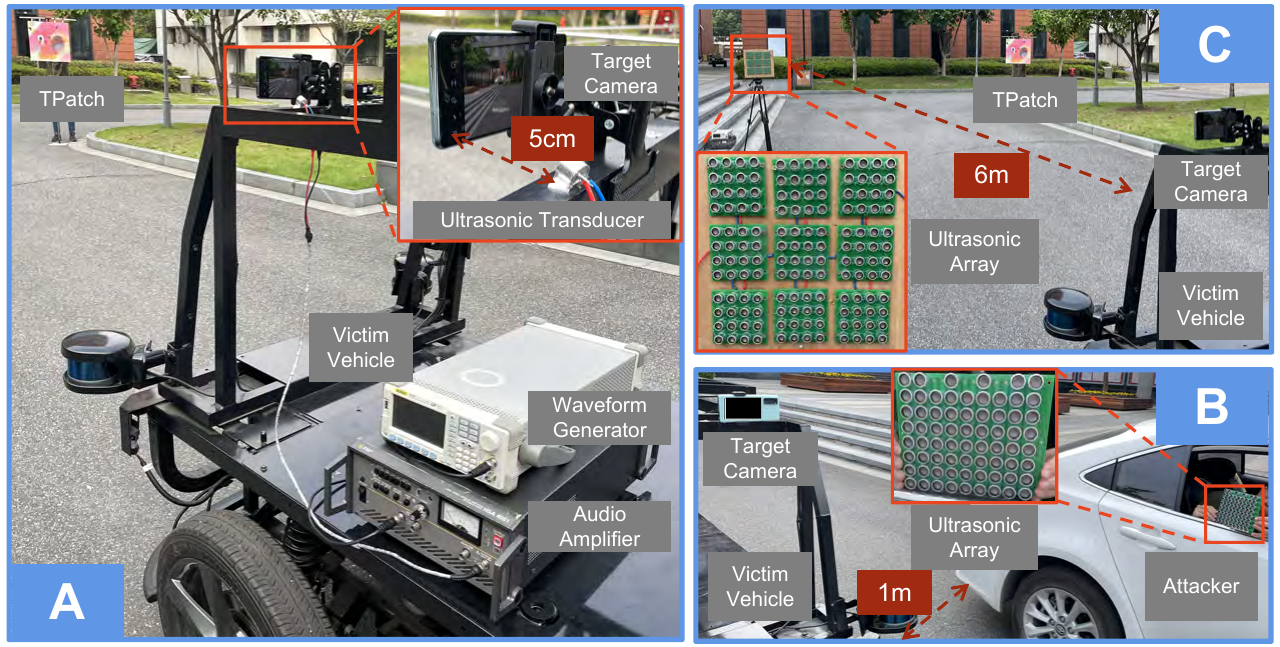}
    \vspace{-0.1in}
    \caption{Experimental setups in three attack scenarios. (A) shows the in-car scenario where a single ultrasonic transducer is attached near the camera to inject acoustic signals. (B) shows the car-to-car scenario where the attacker launches an acoustic injection from the following car. (C) shows the roadside scenario where a stationary ultrasonic array is utilized to conduct roadside attacks.}
    \label{fig:setup}
    \vspace{-0.1in}
\end{figure}

\textbf{Real-world Setup.}
We utilize an advanced Apollo Kit autonomous vehicle as the victim vehicle, a platform to experiment with future self-driving techniques. Its sizes are 2.77 m (length) * 1.01 m (width) * 1.67 m (height). We mount a smartphone, e.g., Samsung S20 or iPhone 7, on the front of the vehicle to mimic its computer vision system. The acoustic signals are generated with an arbitrary waveform generator, an audio amplifier and ultrasonic devices, as shown in Fig.~\ref{fig:setup}. The detailed setup varies with different attack scenarios, i.e., in-car, car-to-car, and roadside scenarios. The required attack power and the scale of ultrasonic devices increase as the attack range rises. We use the same waveform generator and audio amplifier to drive the ultrasonic transducers for convenience. Yet the realistic setup can be on a much smaller scale, especially for the in-car scenario. For HA, the \tp is printed on a sticker attached to a normal stop sign board. For CA, the \tp is printed on a board, with a realistic size of $60cm \times 60cm$. All the driving experiments are done on the closed roads in our institute.

\textbf{Methodology and Metric.}
In the real-world evaluation, we take the in-car scenario as the default scenario, and the attack performance in different scenarios is investigated standalone in Sec.~\ref{subsubsec:imple}. The victim vehicle is driving towards the \tp at a constant speed meanwhile keeping the \tp within the view of camera. The vehicle starts at a distance of round 15 m to the \tp and stops at a distance of 1-2 m to the \tp. The smartphone records the videos of driving experiments, which is recognized by the victim detector on the server. To have a better understanding of the effectiveness of the \tp, we split the positive trigger and negative trigger evaluation and adopt a new metric, i.e., the best attack success rate in n consecutive frames $f_{succ}^{max(n)}$~\cite{zhao2019seeing,shen2022sok}, to evaluate the recognition of videos. The metric is formulated as Eq.~\ref{eq:metric2}:
\begin{equation}
    f_{succ}^{max(n)} = \max_j \frac{1}{n} \sum_{i=1}^{n} 
    \mathbb{I}_{F\left(x_{i+j}\right) = y_a} (x_{i+j})
    \label{eq:metric2}
\end{equation}
where $x_i$ denotes the $i_{th}$ frame of the captured video $x$. To quantify our evaluation, we choose three frame lengths $n$ 50, 100, and 150, i.e., $f_{succ}^{max(50)}$, $f_{succ}^{max(100)}$, and $f_{succ}^{max(150)}$.

\subsubsection{Overall Performance}

To evaluate the overall performance, we craft 6 \tpes (3 for HA and 3 for CA) against three detectors, i.e., Faster R-CNN, YOLO V3, and YOLO V5. The victim vehicle is configured to a slow speed, 5 km/h, and the average duration of captured video clips is around 10 s, and then the number of frames is about 300 at an fps of 30. Since every \tp is trained against a specific detector by gradient-based optimization, we investigate its transferability via recognizing the same video clips with the other black-box detectors.

Tab.~\ref{rw:overall-ca} demonstrate the overall performance of HA and CA. It is indeed possible to achieve transfer attacks in the real world. For CA, the average $f_{succ}^{max(50)}$, $f_{succ}^{max(100)}$, and $f_{succ}^{max(150)}$, are 90.2\%, 79.9\%, and 71.6\%, respectively, where the white-box ones are 100\%, 96.0\%, and 86.4\% while the transferred ones are 85.3\%, 71.8\%, and 64.2\%. For HA, the average of $f_{succ}^{max(150)}$ is 100\% (white-box) and 98.3\% (transferred), which has a better performance than CA. The possible reason is that the HA \tp attached to the normal stop sign destroys the robust feature of the stop sign, which is agnostic to the recognition algorithm, which contributes to the blurring attacks. For comparison, we test the pure blur attack case whereby the stop sign has no patch and the camera is under similar levels of acoustic signal injection and found that the pure blur attack is difficult to hide the stop sign, where $f_{succ}^{max(150)} = 3.3\%$.

\subsubsection{Attack Performance in Various Scenarios}
\label{subsubsec:imple}

The implementation of acoustic signal injection can vary for different attack scenarios, i.e., in-car, car-to-car, and roadside scenarios. We summarize the setups of different attack scenarios in Tab.~\ref{rw:implement2}.

\begin{itemize}
    \item For an in-car scenario, an attacker may place the attack device near the target, whereby the attack device can be one ultrasonic transducer and may be left there during car maintenance. The attack device can be in a small size to be stealthy. 
    \item For a car-to-car scenario, an attacker may drive a car to follow the victim's vehicle and emit acoustic signals with a transducer array. The attack distance is about 1 m.
    \item For a roadside scenario, an attacker may set up an ultrasonic array on the roadside, emitting acoustic signals to a driving-by victim vehicle. The farthest attack distance can be up to 6 m.
\end{itemize}

Note that in the roadside attack scenario, the attacker has to handle the Doppler effects since there may exist a relative speed between the ultrasonic devices and the victim's vehicle. Details of discussion on the Doppler effects are referred to Appendix~\ref{app:doppler}.

\begin{table}[pt]
\caption{Setups of three attack scenarios.}
\label{rw:implement2}
\vspace{-0.1in}
\begin{center}
\centering
\scalebox{0.75}{
\begin{tabular}{c|c|c|c|c}
    \hline
    Scenario   & Non-intrusive & Distance & Device Scale & Power \\ \hline \hline
    in-car     & \ding{55} & 5 cm  & $1.6cm \times 1.6cm$ & 0.08W \\ \hline
    car-to-car & \ding{51} & 1 m   & $15cm \times 15cm$   & 7W    \\ \hline
    roadside   & \ding{51} & 6 m   & $30cm \times 30cm$   & 40W   \\ \hline
\end{tabular}}
\end{center}
\vspace{-0.25in}
\end{table}

To investigate the effectiveness of \tp among various attack scenarios, we conduct experiments for each attack scenario and summarize the results in Tab.~\ref{rw:implement}. For the in-car scenario, the $f_{succ}^{max(150)}$ is 100\% for HA and 96.7\% for CA, respectively. The attacks of an in-car scenario achieved the best performance because the transducer is close to the camera and remains relatively stationary to the camera, which results in imposing stable acoustic signals on the camera. Nevertheless, even with the car-to-car or roadside scenarios whereby the attack transducers are placed external to the victim vehicle, we can still achieve HA with a $f_{succ}^{max(150)}$ of over 93.3\%, and achieve CA with a $f_{succ}^{max(150)}$ of over 86.7\%.

\begin{table}[pt]
\caption{Effectiveness in three attack scenarios.}
\label{rw:implement}
\vspace{-0.1in}
\begin{center}
\centering
\scalebox{0.80}{
\begin{tabular}{c|c|c|c|c}
    \hline
    Attack Type & Scenario   & $f_{succ}^{max(50)}$ 
    & $f_{succ}^{max(100)}$  & $f_{succ}^{max(150)}$ \\ \hline \hline
    HA  & in-car     & 100.0\% & 100.0\% & 100.0\% \\ \hline
    HA  & car-to-car & 100.0\% & 99.0\%  & 96.0\%  \\ \hline
    HA  & roadside   & 100.0\% & 99.0\%  & 93.3\%  \\ \hline
    CA  & in-car     & 100.0\% & 99.0\%  & 96.7\%  \\ \hline
    CA  & car-to-car & 96.0\%  & 91.0\%  & 86.7\%  \\ \hline
    CA  & roadside   & 98.0\%  & 97.0\%  & 91.3\%  \\ \hline
\end{tabular}}
\end{center}
\vspace{-0.25in}
\end{table}

To explore the upper bound of the attack distance, we conduct experiments with various attack distances. We utilize the roadside attack setup with a $30cm*30cm$ ultrasonic transducer array at an apparent power level of 40W. Tab.~\ref{rw:distance} lists the results of $f_{succ}^{max(50)}$ at different attack distances from 2 m to 7 m. We find that we can achieve a $f_{succ}^{max(50)}$ of 100\% for HA with an attack distance up to 6 m, and a $f_{succ}^{max(50)}$ of 98\% for CA with an attack distance up to 5 m. The attack performance begins to decrease when the distance exceeds 5 m, which is because sounds attenuate with distances. Based on the inverse square law, the energy of sound received at 7 m is almost half of that at 5 m. The attenuated acoustic signals are more likely to be affected by environmental factors, leading to a degradation of the attack performance in consecutive frames.

\begin{table}[pt]
\caption{Impact of attack distances.}
\label{rw:distance}
\vspace{-0.05in}
\begin{center}
\centering
\setlength{\tabcolsep}{3mm}{
\scalebox{0.80}{
\begin{tabular}{c|c|c|c|c|c|c}
    \hline
    Distance & 2 m & 3 m & 4 m & 5 m & 6 m & 7 m \\ \hline \hline
    HA & 100\% & 100\% & 100\% & 100\% & 100\% & 70\% \\ \hline 
    CA & 100\% & 100\% & 98\%  & 98\%  & 74\%  & 48\% \\ \hline
\end{tabular}}}
\end{center}
\vspace{-0.2in}
\end{table}

To investigate whether the attacks can affect various cameras, we attack another camera equipped with image stabilization techniques, iPhone7, in addition to Samsung S20. Since iPhone7 mounts a different motion sensor from the one on Samsung S20, their resonant frequencies are different, i.e., 20 kHz for Samsung S20 and 27 kHz for iPhone7. But we can cause a similar image blur on different cameras via acoustic signals with different frequencies. As shown in Tab.~\ref{tab:camera}, we evaluate both HA and CA on iPhone7 and find that our attacks can work with iPhone7 as well with a $f_{succ}^{max(150)}$ of 100\% for HA and a $f_{succ}^{max(150)}$ of 94.0\% for CA. 

\begin{table}[pt]
\caption{Effectiveness with different cameras.}
\label{tab:camera}
\vspace{0.05in}
\centering
\scalebox{0.75}{
\begin{tabular}{c|c|c|c|c|c}
    \hline
    Attack Type & Camera & Freq. & $f_{succ}^{max(50)}$ & $f_{succ}^{max(100)}$ & $f_{succ}^{max(150)}$ \\ \hline \hline
    HA  & Samsung S20 & 20 kHz & 100.0\% & 100.0\% & 100.0\% \\ \hline
    HA  & iPhone7     & 27 kHz & 100.0\% & 100.0\% & 100.0\% \\ \hline
    CA  & Samsung S20 & 20 kHz & 100.0\% & 98.0\%  & 92.7\%  \\ \hline
    CA  & iPhone7     & 27 kHz & 98.0\%  & 93.0\%  & 94.0\%  \\ \hline
\end{tabular}}
\vspace{-0.2in}
\end{table}

\subsubsection{Impact of Environmental Factors}

In this subsection, we investigate how those common environmental factors influence the attack performance of the \tp. We mainly take three aspects of environmental factors into consideration, i.e., car movements, car vibrations, and lighting conditions. We conduct two series of experiments to evaluate the attack performance for positive triggers and the mis-triggering rate for negative triggers.

\begin{figure}[pt]
    \centering
    \includegraphics[width=.98\linewidth]{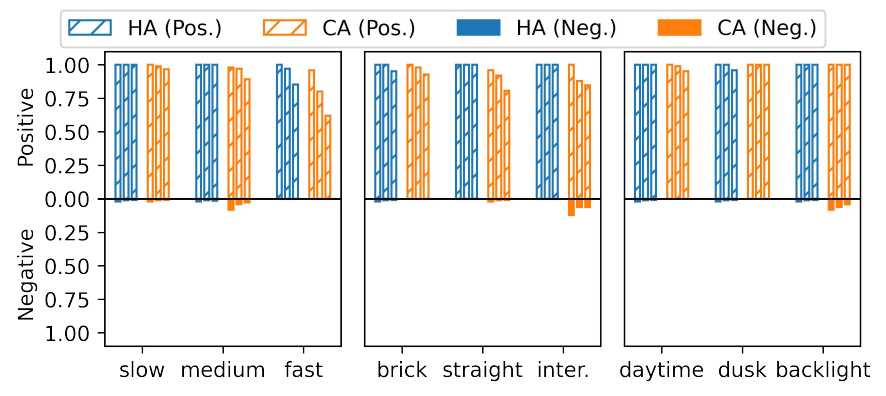}
    \vspace{-0.15in}
    \caption{Impact of environmental factors on HA (blur bars) and CA (orange bars), including the speeds of vehicle (left), the road types (middle), and the lighting conditions (right). The three adjacent bars denote $f_{succ}^{max(50)}$, $f_{succ}^{max(100)}$, and $f_{succ}^{max(150)}$.}
    \label{fig:scenario}
    \vspace{-0.1in}
\end{figure}

\textbf{Car Movement.}
For car movements, we evaluated their impact by conducting experiments with three car speeds: (1) slow (5 km/h), (2) medium (10 km/h), and (3) fast (15 km/h). We limited the highest experimental speed to 15 km/h for safety reasons. As shown in Fig.~\ref{fig:scenario}, we can achieve a $f_{succ}^{max(50)}$ of 100\% for HA and a $f_{succ}^{max(50)}$ of 96\% for CA with any car speed. The $f_{succ}^{max(150)}$ of HA and CA at high speed are lower than the rest because the number of captured frames decreases when the vehicle drives fast. On the other hand, the \tp is hardly mis-triggered as the car moves, and the average mis-triggering rate is 0.6\% for HA and 1.1\% for CA, w.r.t. $f_{succ}^{max(150)}$. 

\textbf{Car Vibration.}
For car vibrations, we conduct experiments on three roads with various conditions: (1) an even brick road, (2) a straight asphalt road, and (3) an asphalt intersection. We find that the impact of car vibrations is abrupt. Most of the time, the car is in relatively stable driving conditions. As shown in Fig.~\ref{fig:scenario}, the \tp has a bit better performance on the brick road than the other roads. The car vibration is possible to trigger our \tp as long as the motion blur is similar to the positive trigger, but it is occasional and has a limited mis-triggering rate, 0.2\% for HA and 2.2\% for CA, w.r.t. $f_{succ}^{max(150)}$. 

\textbf{Lighting condition.}
Lighting conditions may change the color of the captured images and thus influence the effectiveness of \tp. We investigate three lighting conditions, e.g., daytime (2000 lux), dusk (400 lux), and backlight (100 lux). As shown in Fig.~\ref{fig:scenario}, the \tp is robust to different lighting conditions and achieves nearly 100\% in all situations. It is because we have considered the possible color shifts in the EoT training of \tp, including brightness, contrast, saturation, and hue.

\section{Countermeasures and Limitation}

\subsection{Countermeasures}

\begin{figure*}[pt]
\centering
\includegraphics[width=.95\linewidth]{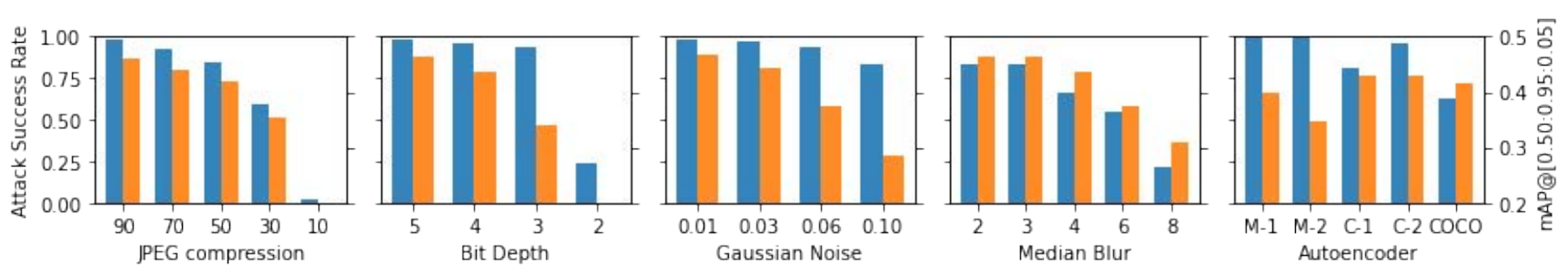}
\label{fig:new-def}
\vspace{-0.2in}
\caption{
ASRs (blue bars) and mAPs in benign scenarios (orange bars) with the input-transformed defenses. The x-axis represents the strength of defenses for all defenses except the denoise autoencoders. The character of 'M' and 'C' denotes the MNIST and CIFAR-10, respectively, as the training dataset of autoencoders. The number '1' and '2' denotes the two different model architecture proposed in~\cite{meng2017magnet}.
}
\vspace{-0.1in}
\end{figure*}

In this section, we discuss four potential defenses against \tp from three levels, i.e., physical signal protection (sensor level), input-transformed defenses (algorithm level), model robustness improvement (algorithm level), and sensor fusion techniques (system level).

\textbf{Physical Signal Protection.} 
In this paper, we exploit the image blur caused by acoustic signals as the trigger of attacks, which is enabled by the vulnerability of MEMS inertial sensors to acoustic injection attacks. Microfiber metallic fabrics or MEMS fabricated acoustic metamaterials can be used to isolate the attacking sound. We can also design a secure low-pass filter to eliminate out-of-band analog signals, thus inhibiting the ability of the adversary to control the sensor output through signal aliasing~\cite{trippel2017walnut}. 

\textbf{Input-Transformed Defenses.}
A series of works intended to exploit the instability of adversarial examples to detect or mitigate this type of attack, typically via transformations on the input images, e.g., JPEG compression~\cite{dziugaite2016study}, bit depth reduction~\cite{xu2017feature}, Gaussian noise~\cite{zhang2019defending}, median filter~\cite{xu2017feature}, and denoise autoencoder~\cite{meng2017magnet}.
Due to their simplicity, these defenses are evaluated in recent relevant work like~\cite{sato2021dirty} and~\cite{cao2021invisible}.
We use the ASR to evaluate the effectiveness of the defense and use the mean Average-Precision (mAP) to evaluate its impact on the performance of the detection model.
As shown in Fig.~\ref{fig:new-def}, we find that with a large defense strength, the tested methods can defend our attacks to some extent but impair the performance of the detection model.
For safety-related applications like self-driving, such performance drops of recognition may not be acceptable. Therefore, using the input-transformed defenses cannot easily defeat our attacks.

\textbf{Model Robustness Improvement.}
The key factor that contributes to trigger-ability is the vulnerability of recognition models to the distortion. We can boost the resistance of the recognition model in two ways: (1) train a more robust model that tolerates the image distortion to an extent, and (2) detect the distortion via evaluating the image quality by metrics, e.g., PSNR and MS-SSIM with the last frame, and then, lower the importance of the recognition results of low-quality images. 

\textbf{Sensor Fusion Techniques.}
The general ways to cope with the abnormal situations that exist in the scenario of autonomous driving still work with \tp, e.g., the fusion of 3d point cloud collected by Lidar and visual perception of the camera in obstacle detection, the reference of the HD maps that contain rich information about traffic signs, etc.

\subsection{Limitation}

\tp attacks have the following limitations at present. First, our evaluation mainly focuses on studying the effects of \tp attacks on the AI component level rather than the AV system level~\cite{shen2022sok}. A system-level evaluation will help better understand the real-world impacts of our attacks. We will have a closed-loop simulation with the other self-driving components, e.g., prediction, planning, and control for evaluation in the future. Second, we only investigate the acoustic signal attacks as the trigger of \tp. More evaluation on the other sensor attacks is demanded. Third, although we have enhanced the visual camouflage of \tp with the content loss, it may still alter acute victims. Further improvement on the camouflage of \tp is required. Fourth, the attack device for long range attacks can be noticeable at present. To improve its visual stealthiness, we can (1) employ a more compact array arrangement and (2) fit it with the surrounding environment. For instance, the scale of the attack device in the roadside scenario can be reduced into $20cm \times 20cm$ by simply eliminating gaps between the transducers. In addition, the adversary could spray the ultrasonic array in a color conform to the surroundings, e.g., green for the vegetation. We designate the aforementioned issues as our future work.

\section{Related Work}

In this section, we summarize the existing works on adversarial attacks from three aspects, i.e., digital noise attacks, physical adversarial attacks, and adversarial sensor attacks.

\textbf{Digital Noise Attacks.} 
It has been heavily studied how to generate adversarial examples against image classifiers by adding invisible noises, e.g., FGSM~\cite{goodfellow2014explaining}, PGD~\cite{madry2017towards}, I-FGSM~\cite{kurakin2016adversarial}, MI-FGSM~\cite{dong2018boosting} and CW attacks~\cite{carlini2017towards}. Other than image classification, Xie et al.~\cite{xie2017adversarial} extend adversarial attacks to object detection and image segmentation.

\textbf{Physical Adversarial Attacks.} 
A bunch of works aim to explore the physical adversarial attacks. Kurakin et al.~\cite{kurakin2016adversarial} first prove the feasibility of printed adversarial attacks. However, the subtle pixel modification limits the attack capability. Adversarial patch~\cite{brown2017adversarial} replaces the additive noise with the placement of the patch. In~\cite{athalye2018synthesizing}, Expectation over Transformation (EoT) is proposed to boost the robustness of adversarial examples. Chen et al.~\cite{chen2018shapeshifter} propose a robust attack, ShapeShifter, to Faster R-CNN in the real world. Song et al.~\cite{song2018physical} successfully attack both Faster R-CNN and YOLO V2. Zhao et al.~\cite{zhao2019seeing} realize long-range attacks by adversarial example nested in the patch. Lovisotto et al.~\cite{lovisotto2021slap} render the stop sign with visible light to realize the adversarial attacks.

\textbf{Adversarial Sensor Attacks.}
There exists another branch of researches that launch sensor attacks on the camera to achieve the adversarial goals. Yan et al.~\cite{yan2016can} demonstrate that the direct laser projection can make the camera blind or even destroyed. Man et al.~\cite{man2020ghostimage} utilize the lens flare to manipulate the recognition of traffic signs. Sayles et al.~\cite{Sayles2021InvisiblePP} propose an optimization of color strips induced by the rolling shutter effect to mislead the image classifier. Ji et al.~\cite{Ji2021PoltergeistAA} propose adversarial blur attacks on object detectors by emitting acoustic signals to hijack the image stabilization system.

\section{Conclusion}

In this paper, we present \tp, a physical adversarial patch triggered by specific physical signals to realize hiding, creating and altering attacks against the vision-based perception module in a targeted autonomous vehicle. Evaluations with three object detectors, YOLO V3/V5 and Faster R-CNN, and eight image classifiers demonstrate the effectiveness of \tp in both the simulation and the real world. This work serves as the first attempt at the adversarial patches triggered by physical signals. Further directions include investigating other signals in addition to the acoustic ones as the triggers.

\section*{Acknowledgement}

\vspace{-0.1in}
We thank our shepherd and the other anonymous authors for their valuable comments and suggestions. We would like to thank our colleagues Bo Yang and Zizhi Jin for help with the driving experiments; Kai Wang for useful discussions. This paper is supported by China NSFC Grant 62222114, 61925109, 62071428, 62271280 and China Postdoctoral Science Foundation Grant BX2021158.
\vspace{-0.1in}

\bibliographystyle{plain}
\bibliography{reference.bib}

\appendix

\section{Design Details of Acoustic Signal Injection}

\subsection{Challenge}
\label{app:challenge}

Previous work~\cite{trippel2017walnut} has demonstrated that injecting acoustic signals into a MEMS motion sensor can induce a constant bias when the signal frequency is an integer multiple of the sampling frequency. Based on it, the attacker can form any signal through amplification or phase modulation. However, this method requires the attacker to monitor the sensor readings of the victim’s car in real-time to synchronize the attack signals, which can be difficult to fulfill in real driving scenarios. Without signal synchronization, the sensor output may suffer from phase shifting and frequency shifting, degrading the performance of manipulation. In the following, we use the one-dimension sensor data as an example to demonstrate their effects.

\begin{figure}[pb]
    \centering
    \includegraphics[width=.98\linewidth]{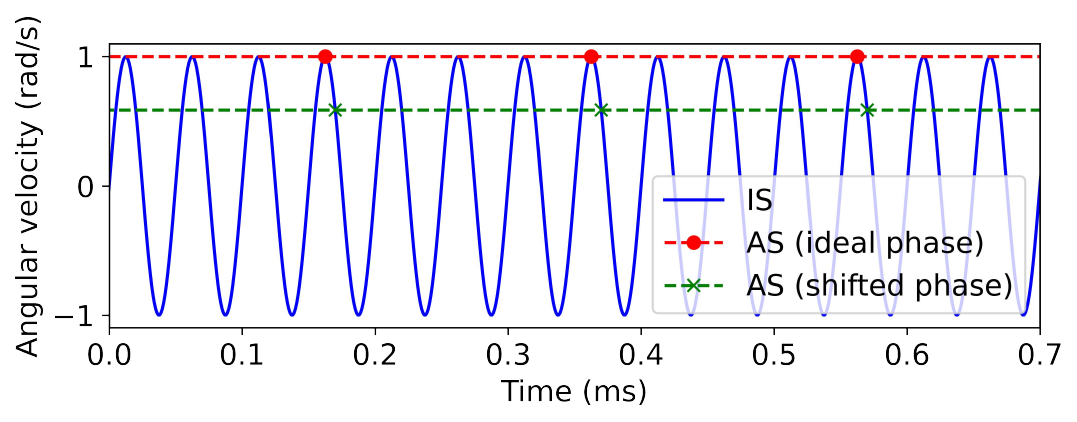}
    \caption{Phase shifting effects on the aliased signals. The blue line denotes the resonance of gyroscope induced by injected signals (IS). The red and green dot lines represent the aliased signals (AS) with different phase.}
    \label{fig:alias-phase}
\end{figure}

\textbf{Phase shifting.} Without synchronization, a random phase shift can be introduced during the aliasing process, as shown in Fig.~\ref{fig:alias-phase}. The random phase shift then introduces a random sensor output, and thus the motion blur with a random strength.

\begin{figure}[pt]
    \centering
    \includegraphics[width=.98\linewidth]{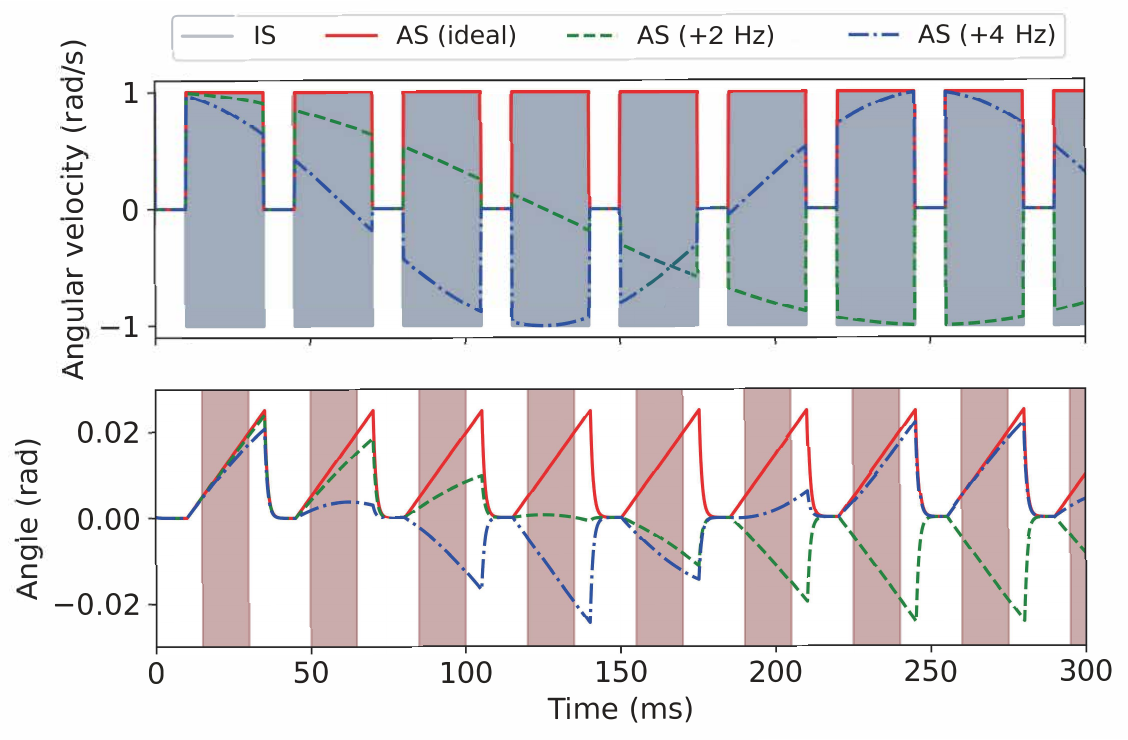}
    \caption{Frequency shifting effects on the aliased signals. The bottom figure shows the moving angles of camera which is the integral of angular velocity shown in the upper figure. The red line represents the ideal situation where the sampling frequency is constant, while the green and blue dot lines represent the consequences caused by different frequency shifts.}
    \label{fig:alias-freq}
\end{figure}

\textbf{Frequency shifting.} Similarly, when without synchronization, the sampling frequency of the MEMS sensor may drift slightly from time to time due to the temperature changes. 
As shown in Fig.~\ref{fig:alias-freq}, a small shift in frequency, e.g., 2 Hz or 4 Hz, can render the camera motion different from the ideal one and varied in continuous frames.


\subsection{Our Approach}
\label{app:approach}
To address the aforementioned issues, we first model the image blur effect. 
A motion-blurred image can be viewed as the superposition of a series of translated clean images, which can be formulated in Eq.~\ref{eq:blur}:
\begin{equation}
L(i,j,t) = \int_{t}^{t+t_e}g(i+\Delta i(\tau),j+\Delta j(\tau))d\tau
\label{eq:blur}
\end{equation}
where $L(i,j,t)$ denotes the $j_{th}$ row and $i_{th}$ column of the image $L$ captured at the moment $t$. $t_e$ is the exposure duration of the camera. $\Delta i(t)$ and $\Delta j(t)$ denote the translation of the X-axis and Y-axis of the image at the moment $t$. $g(i,j)$ denotes the $j_{th}$ row and $i_{th}$ column of the stationary image $g$.

Based on this formulation, we find the key to reducing the impact of phase/frequency shifting is to make the image blur independent of the time $t$.
To achieve it, we propose to shift the frequency of the injected signal from the integer multiple of the sampling frequency $n f_s$ to the frequency $n f_s \pm f_e$, where $f_e$ represents the exposure frequency. In this way, the frequency of the aliased signal 
can be given with $f_a = |f - n f_s| = |n f_s \pm f_e - n f_s| = f_e$, which equals the exposure frequency. Since the aliased signal determines the sensor output and thus the camera motion, the translation functions of the X-axis $\Delta i(t)$ and Y-axis $\Delta j(t)$ share the same period with the camera exposure duration $t_e$, as formulated in Eq.~\ref{eq:period}:
\begin{equation}
\Delta i(t) = \Delta i(t+t_e),\quad \Delta j(t) = \Delta j(t+t_e),\quad \forall t
\label{eq:period}
\end{equation}
In this way, we can inject a stable blur pattern that is time-independent as shown in Eq.~\ref{eq:stable}, which eliminates the impact of phase shifting.
\begin{equation}
L(i,j,t_1) = L(i,j,t_2), \quad \forall t_1, t_2
\label{eq:stable}
\end{equation}
The proof of the property is provided as follows:
\begin{scriptsize}
\begin{equation}
\nonumber
\begin{aligned}
L(i,j,t_1) &= \int_{t_1}^{t_1+t_e} g(i+\Delta i(\tau), j+\Delta j(\tau)) d\tau \\
&= \int_{t_1}^{t_2} g(i+\Delta i(\tau), j+\Delta j(\tau)) d\tau 
+ \int_{t_2}^{t_1+t_e} g(i+\Delta i(\tau), j+\Delta j(\tau)) d\tau \\
&= \int_{t_1}^{t_2} g(i+\Delta i(\tau+t_e), j+\Delta j(\tau+t_e)) d\tau 
+ \int_{t_2}^{t_1+t_e} g(i+\Delta i(\tau), j+\Delta j(\tau)) d\tau \\
&= \int_{t_1+t_e}^{t_2+t_e} g(i+\Delta i(\tau), j+\Delta j(\tau)) d\tau 
+ \int_{t_2}^{t_1+t_e} g(i+\Delta i(\tau), j+\Delta j(\tau)) d\tau \\
&= \int_{t_2}^{t_2+t_e} g(i+\Delta i(\tau), j+\Delta j(\tau)) d\tau = L(i,j,t_2)
\end{aligned}
\end{equation}
\end{scriptsize}

\begin{figure}[pt]
\centering
\includegraphics[width=.98\linewidth]{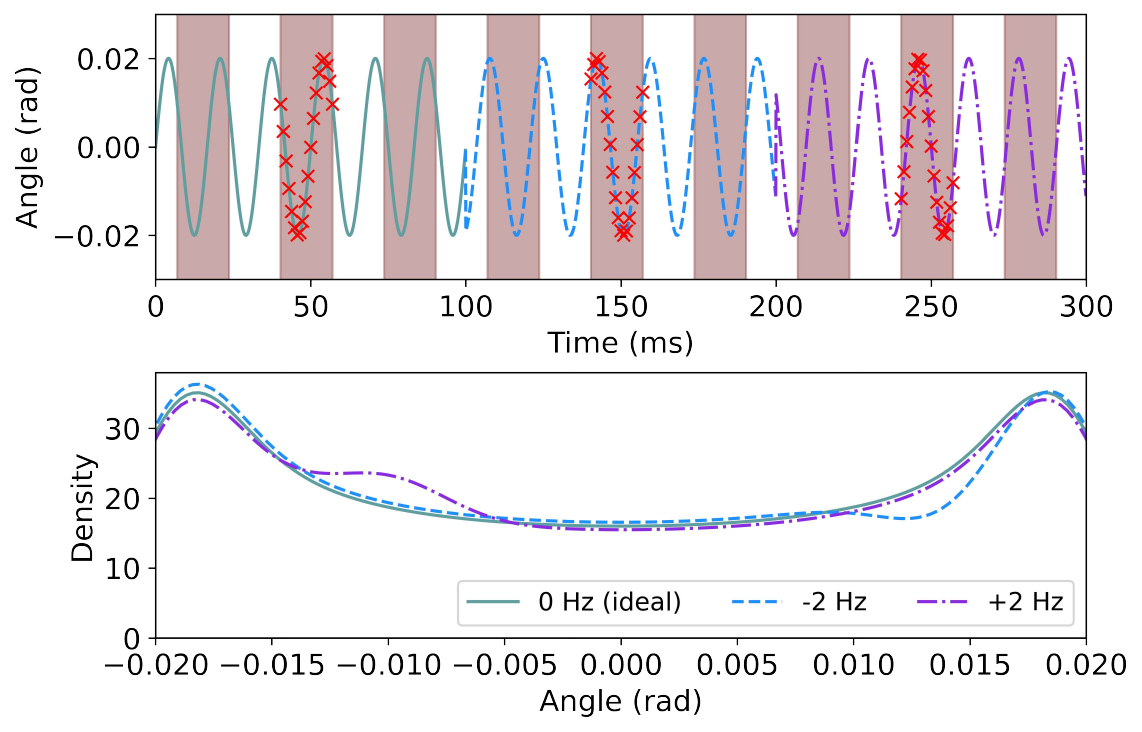}
\caption{Aliased signals with frequency shifting. The upper figure shows the image translation caused by aliased sine signals, where the curves with different colors represent different frequency shifts and the red shadows represent the exposure duration. The bottom figure shows the density of image translation.}
\label{fig:alias-sin}
\end{figure}

For frequency shifting, our approach cannot eliminate it but can reduce its impact. 
The reason is that the aliased signal generated by our approach has a much higher frequency than the frequency shift caused by the temperature drift, making its impact not significant.
For instance, given a desired aliased signal of 60 Hz and a frequency shifting of $k=2$ Hz,  the actual aliased signal is 58 Hz or 62 Hz, which still has a high overlap with the idea one, as shown in Fig.~\ref{fig:alias-sin}
 

\section{Impact of Doppler Effects} \label{app:doppler}

In the roadside attack scenario, a relative speed can exist between the driving-by vehicle and the stationary attack device, resulting in a frequency shift of the injected signals due to the Doppler effects. Given a relative speed of 1 m/s and an emitted signal frequency of 20 kHz, the frequency shift of injected acoustic signal is 58.8 Hz. To address it, the attacker can compensate for the frequency shifting in real-time by externally measuring the speed of the vehicle via a certain common approach, e.g., LiDAR~\cite{schwemmer1993doppler}. For the reason of engineering, we configure the car at a constant speed to simulate the scenario where the adversary has obtained the vehicle speed from real-time measurements.



\section{Explainable-AI based Detection Techniques}

\begin{figure}[ptb]
    \centering
    \includegraphics[width=.95\linewidth]{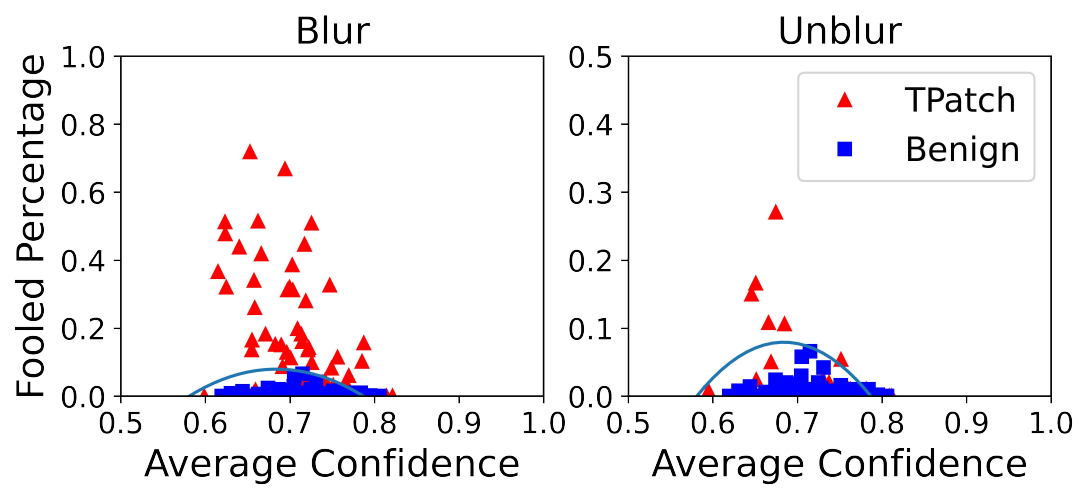}
    \caption{Visualization of SentiNet results. The red triangles and blue squares respectively denote \tp and benign examples. The curve represents the decision boundary.}
    \label{fig:sentinet}
\end{figure}

Recently, a few approaches are proposed to defend the attack of adversarial patches, e.g., \textit{SentiNet}~\cite{chou2020sentinet}, \textit{PatchGuard}~\cite{xiang2021patchguard}, etc. The common methodology among existing defenses is to utilize the explainable-AI techniques, e.g., Grad-CAM~\cite{selvaraju2017grad} in \textit{SentiNet} and BagNet~\cite{brendel2018approximating} in \textit{PatchGuard}, to localize the potential adversarial patch, whose saliency is abnormally high. Once the \tp is triggered, it would work like the normal adversarial patch; hence, it is feasible to tackle our attack with similar approaches. We report the defense results of \textit{SentiNet} in Fig.~\ref{fig:sentinet}, where the accuracy of \textit{SentiNet} is relatively low when \tp is untriggered. We suppose that there are pilot systems that are equipped with these state-of-the-art detection techniques. We want to investigate whether it is possible to detect the \tp via feature analysis.

We use Grad-CAM and BagNet to extract the feature saliency map of the target class. Here we make a strong assumption that the defender knows the target class of attack. If the defender is still unable to detect the presence of the \tp despite having additional prior conditions, it means that this detection method is also difficult to succeed with zero prior. We follow the methodology of original papers to process the saliency map. In \textit{SentiNet}, the defender uses a threshold to generate the binary mask and then crops the masked part of the image, which potentially contains the adversarial patch. We keep 30\% of the pixels with the highest saliency for evaluation. In \textit{PatchGuard}, the defender uses BagNet as the recognition model, whose final convolutional layer can generate a local feature map of each class, and utilizes a sliding window to detect the most salient area. We design the \tp as $48\times48$ pixels and the corresponding best window size is $6\times6$. After finding the most salient window, we can map the window in the feature space to the mask in the image pixel space.

\begin{figure}[ptb]
    \centering
    \includegraphics[width=.95\linewidth]{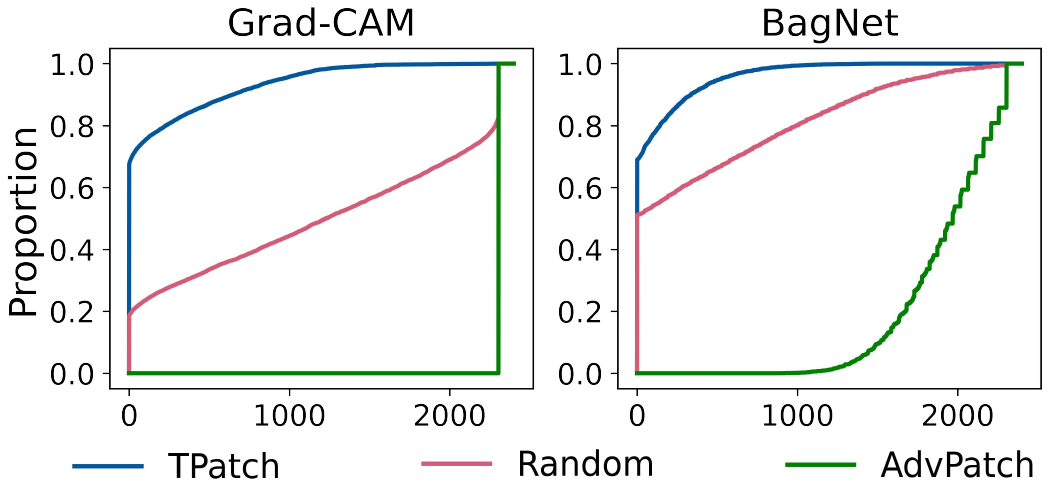}
    \caption{CDF of the number of masked pixels. We use ResNet-50 and BagNet-33 as the recognition model respectively for the experiments of Grad-CAM and BagNet.}
    \label{fig:stealth}
\end{figure}

We define the masked pixel numbers of the \tp as the metric of evaluation, where zero stands for the complete failure of detection and the total pixel number of the patch stands for the success of detection. We conduct experiments on three types of patches, i.e., random patch, normal adversarial patch, and our \tp, and plot the overall result with the format of cumulative distribution function (CDF). As shown in Fig.~\ref{fig:stealth}, both methods have an excellent performance in detecting the normal adversarial patch. However, both methods perform badly on detecting the \tp and never mask the whole patch. Surprisingly, the performance of detecting the \tp is even worse than the one of detecting the random patch, which represents the probability of accidentally masking the irrelevant objects. 

\section{Physical Experiments on Altering Attacks}

We conduct indoor physical experiments to investigate the effectiveness of altering attacks on classifiers. We place the \tp with a mouse (an existed class in ImageNet~\cite{deng2009imagenet}) in the view of camera and use a switch to control the acoustic signal injection. When the switch is off, the classifiers correctly recognize the mouse, and the target class, traffic light, is not within the top-5 recognition results. When the switch is on, the top-1 recognition result is altered into the traffic light. All the \tpes achieve $f_{succ}^{max(150)}$ of 100\% on three tested classifiers, i.e., VGG16, Res50, and Incv3.

\section{Impact of Other Acoustic Signals}

We conduct experiments with three representative types of acoustic signals: (1) in-car music, (2) road noises, and (3) traffic horn sounds. During attacks, we played each of them at 110dB using a speaker at a distance of 5cm from the camera. From the results, we find that none of them has effects on either the captured images or the motion sensor readings. This is because common background noises are in a lower frequency band (<10kHz) than the resonant frequencies of the motion sensors tested in our experiments, i.e., higher than 20kHz.

\section{Impact of Phase Differences}
\label{app:phase}

\begin{figure}[h]
    \centering
    \includegraphics[width=.95\linewidth]{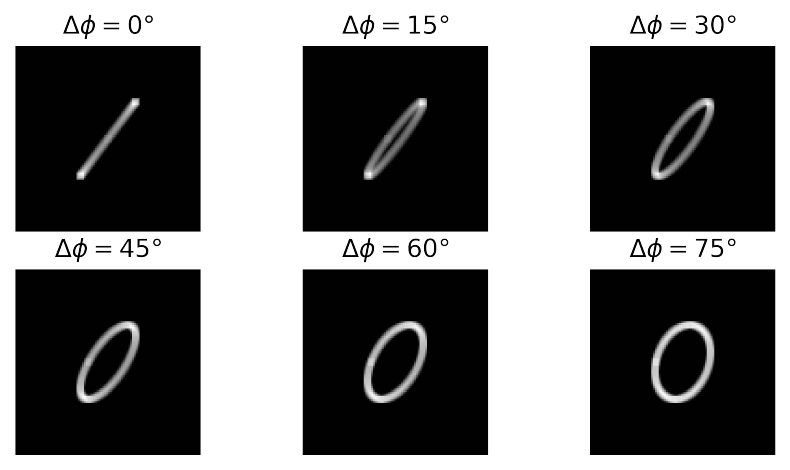}
    \vspace{-0.2in}
    \caption{PSF kernels of different phase differences.}\
    \label{fig:phase-diff2}
    \vspace{-0.2in}
\end{figure}
\begin{figure}[h]
    \centering
    \includegraphics[width=.9\linewidth]{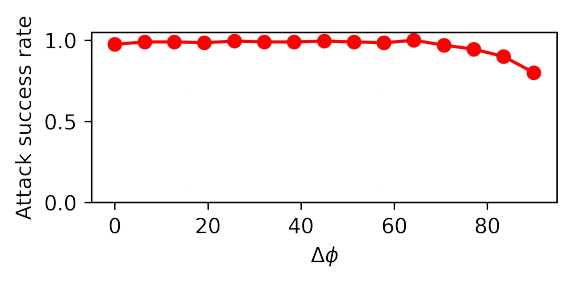}
    \vspace{-0.2in}
    \caption{ASRs with different phase differences.}
    \label{fig:phase-diff}
\end{figure}

With the same strength $A$ and orientation $\psi$ of motion blur, different phase differences can make the PSF kernel various shapes, as illustrated in Fig.~\ref{fig:phase-diff2}, while the phase differences have few impacts on the ASRs of \tp, as shown in Fig.~\ref{fig:phase-diff}.

\end{document}